\documentclass[twocolumn,aps,prl,superscriptaddress,english]{revtex4-2}

\usepackage[colorlinks=true,allcolors=blue]{hyperref}
\usepackage{amsmath}
\usepackage{amssymb}
\usepackage{graphicx}
\usepackage{babel}
\usepackage{braket}
\usepackage{mathrsfs}
\usepackage{amsfonts}
\usepackage{epstopdf}
\usepackage{multirow}
\usepackage{color}
\usepackage{bm}
\usepackage{amsthm}

\usepackage{xcolor}
\usepackage{tikz,fp}
\usepackage{tikz-cd}
\usepackage{adjustbox}
\usepackage{enumitem}
\usetikzlibrary{arrows}
\usetikzlibrary{intersections}
\usetikzlibrary{shapes.geometric}
\usetikzlibrary{decorations.pathreplacing}
\usetikzlibrary{decorations.pathmorphing, patterns,shapes,fixedpointarithmetic}
\usetikzlibrary{decorations.markings}
\usetikzlibrary{snakes}

\tikzset{
  mid arrow/.style={postaction={decorate,decoration={
        markings,
        mark=at position .575 with {\arrow[#1]{stealth}}
      }}},
  near arrow/.style={postaction={decorate,decoration={
        markings,
        mark=at position .275 with {\arrow[#1]{stealth}}
      }}},
   far arrow/.style={postaction={decorate,decoration={
        markings,
        mark=at position .800 with {\arrow[#1]{stealth}}
      }}},
}
\tikzset{snake it/.style={decorate, decoration=snake}}

\newcommand{\ee}{\mathrm{e}} 
\newcommand{\ii}{\operatorname{i}} 
\newcommand{\HH}{\mathcal{H}} 
\newcommand{\KK}{\mathcal{K}} 
\newcommand{\ZZ}{\mathbb{Z}} 

\newcommand{\EE}{\mathcal{E}} 

\newcommand{\Tr}{\mathrm{Tr}}
\newcommand{\dg}{\dagger}

\newcommand{\wt}[1]{\widetilde{#1}}
\newcommand{\vbra}[1]{\left(#1\right|}
\newcommand{\vket}[1]{\left|#1\right)}
\newcommand{\vbraket}[2]{\left(#1\middle|#2\right)}
\newcommand{\vbraopket}[3]{\left(#1\middle|#2\middle|#3\right)}
\newcommand{\kket}[1]{\lvert#1\rangle\!\rangle}
\newcommand{\bbra}[1]{\langle\!\langle#1\rvert}
\newcommand{\lrangle}[1]{\langle#1\rangle}
\newcommand{\abs}[1]{\left\lvert#1\right\rvert}

\newcommand{\todo}[1]{{\color{brown}TODO: #1}}

\newcommand{\YZ}[1]{{\color[rgb]{0.13,0.55,0.13} [YZ: #1]}}

\makeatletter

\newcommand*{\wideboxed}[1]{\setlength{\fboxsep}{1ex}%
  \fbox{\m@th$\displaystyle#1$}}
\makeatother

\makeatletter
\def\maketitle{
\@author@finish
\title@column\titleblock@produce
\suppressfloats[t]
}
\makeatother
\begin{document}

\title{From Topological Order to Mixed-State Phases: A Ground-State Probe of Fractionalized Excitations}
\author{Yunlong Zang}
\thanks{These authors contributed equally.}
\affiliation{Kavli Institute for Theoretical Sciences and School of Quantum,\\ University of Chinese Academy of Sciences, Beijing 100190, China}
\author{Yu-Bin Li}
\thanks{These authors contributed equally.}
\affiliation{Kavli Institute for Theoretical Sciences and School of Quantum,\\ University of Chinese Academy of Sciences, Beijing 100190, China}
\author{Shenghan Jiang}
\email{jiangsh@ucas.ac.cn}
\affiliation{Kavli Institute for Theoretical Sciences and School of Quantum,\\ University of Chinese Academy of Sciences, Beijing 100190, China}
\date{\today}

\begin{abstract}
    How do we detect topological phases from a single ground state?
    Entanglement entropy and spectrum have long been the standard tools — but the reduced density matrix~(RDM) itself contains far more information.
    We show that the RDM of a 2D topologically ordered system, expressed at the entanglement cut, realizes a 1D mixed-state phase.
    For the $\ZZ_2$ toric code phase, it is a 1D $\ZZ_2$ strong-to-weak spontaneous symmetry breaking~(SW-SSB) phase, where deconfinement of anyons manifests as the short-range correlation of both $\ZZ_2$ charge and $\ZZ_2$ domain-wall in the RDM.
    The bulk $e$-$m$ duality translates into a Kramers--Wannier self-duality of the SW-SSB phase.
    Extending the framework to gapped $\ZZ_2$ spin liquids, the global spin-rotation symmetry manifests as an additional weak symmetry for the 1D RDM.
    Spin-$\frac{1}{2}$ spinons result in a cusp on the disorder parameter of spin-rotation at $\theta=\pi$, providing a direct, ground-state signature of symmetry fractionalization.
    We verify this prediction analytically using the matrix product density operator formalism and numerically for the kagome-lattice resonating valence bond state.
    The proposed observable requires only a single ground-state wavefunction, making it amenable to quantum simulation platforms.
\end{abstract}

\maketitle

\emph{Introduction.}
Topologically ordered phases, characterized by long-range entanglement and anyonic excitations, stand as a cornerstone of modern condensed matter physics~\cite{Wen1990a,kitaev2003fault,kitaev2006anyons,Wenreview2017}.
While routinely identified by global signatures such as ground-state degeneracy on the torus~\cite{Wen1990}, their experimental detection remains a formidable challenge — one that demands new theoretical tools beyond conventional order parameters.

Entanglement-based diagnostics have partially filled this gap.
The topological entanglement entropy (TEE) — a universal subleading constant in the area law — quantifies the total quantum dimension of anyons~\cite{Kitaev2006TEE,Levin2006detecting}.
The entanglement spectrum mimics edge-mode dispersions, particularly for chiral phases~\cite{Li2008entanglement}.
These measures are derived from the reduced density matrix~(RDM), suggesting that the RDM itself should encode richer information about the underlying topological order\cite{Cirac2011,Schuch2013transfer}.

The key insight of this work is that the RDM of a 2D topological phase, evaluated at the entanglement cut, realizes a nontrivial \emph{one-dimensional} mixed state phase~\cite{deGroot2022symmetryprotected,MaWang2023ASPT,lee2023quantum,chen2024symmetry-enforced,sang2024mixed,maTurzillo2025spt,MaZhangetal2025iASPT,Lessa2025SWSSB,sang2025mixedstatephaseslocalreversibility}.
This \emph{pure--mixed correspondence} — recently explored from complementary perspectives~\cite{Luo2025,SchaferNameki2025,Qi2025,Lu2025,sala2025entanglement,xu2025diagnosing,ma2025measurement} — opens a new window into topological order through the lens of mixed-state phases.
Specifically, for the $\ZZ_2$ toric code topological order, we demonstrate that its RDM realizes a 1D $\ZZ_2$ strong-to-weak spontaneous symmetry breaking~(SW-SSB) phase~\cite{Lessa2025SWSSB,lee2023quantum,sala2024spontaneous,lu2024bilayer,Liu2025diagnosingSWSSB,weinstein2024swssb,divi2026local,zhang2026local,liu2026local}, in which a strong symmetry ($\rho=U\rho = \rho U$)\cite{albert2014symmetries,deGroot2022symmetryprotected} is spontaneously broken to a weak one ($\rho=U\rho U^\dagger$).
    Here, the strong $\ZZ_2$ symmetry inherited from the bulk one-form symmetry~\cite{gaiotto2015generalized,JiWen2020} is spontaneously broken to a weak one, and this SW-SSB directly signals the deconfinement of anyons.

This correspondence is powerful in two complementary ways.
First, it inspires a definition of 1D SW-SSB phases based solely on \emph{linear correlators}: the charged-operator correlator and the domain wall correlator~\cite{Kadanoff1971disorder,levin2020constraints} both decay exponentially, manifesting the deconfinement of $e$ and $m$ anyons~\footnote{After the completion of this work, we became aware that a similar definition based on the domain-wall correlator was proposed by Ma and Turzillo~\cite{maTurzillo2025spt}.}.
The bulk $e$-$m$ electromagnetic duality translates into the Kramers-Wannier self-duality of 1D SW-SSB phase\cite{Kramers1941,Li2026}.
Second, the framework, when extended to gapped $\ZZ_2$ spin liquids~\cite{anderson1973,sachdev1992kagome,wen2002quantum,lee2023quantum,Savary2017}, leads to a powerful new method for detecting symmetry fractionalization~\cite{Barkeshli2019SF, Chen2017SF, Luyuanming2024}.
We prove analytically — using a matrix product density operator (MPDO) description of the 1D RDM — that the spin-rotation disorder parameter $\lrangle{U_A(\theta)}$~\cite{wang2024distinguishing,chenTopologicalDisorderParameter2022,huang2022topological,cai2024,liu2014,jiang2023,cai2024disorder,mao2025probing,huang2025mixed,huang2025interaction} develops a \emph{non-analytic cusp} at rotation angle $\theta = \pi$ when $e$-anyon carries spin-$\frac{1}{2}$.
We verify this prediction numerically using the nearest-neighbor resonating valence bond (NN-RVB) state on the kagome lattice.
This provides a direct, ground-state probe of spinons that requires only a single wavefunction, which is readily implementable on quantum simulation platforms.

\emph{RDM of the Toric Code Phase.}
We start by deriving the RDM as a one-dimensional mixed state for the toric code model on a square lattice~\cite{Ho2015}.
As shown in Fig.~\ref{fig:cut}, the system is partitioned into two simply connected regions, $A$ and $\bar A$, with cut length $L$.
The toric code Hamiltonian naturally decomposes as
\begin{equation}
        H=-\sum_{p}\prod_{i\in p}Z_i-\sum_{v}\prod_{i\in v}X_i
        \equiv H_A+H_{\bar A}+H_{\partial A\cup\partial \bar A}
        \label{eq:toric_code_decomp}
\end{equation}
where $H_A$ and $H_{\bar A}$ contain all local terms entirely within $A$ and $\bar{A}$, respectively, and $H_{\partial A\cup\partial \bar A}$ consists of the $2L$ stabilizer terms that intersect the entanglement cut (colored dots in Fig.~\ref{fig:cut}).

As all local terms commute, ground states of $H$ are the simultaneous ground states of all three parts.
We first examine the ground state subspace of $H_A$, denoted as $\HH_{\partial A}$.
At first glance, $\HH_{\partial A}$ contains $L$ local zero-modes at $\partial A$, described by $L$ pairs of Pauli operators commuting with $H_A$:
\begin{equation}
    \tau^{x}_{k}\equiv  X^A_{2k}~,\quad
    \tau^{z}_k\tau^{z}_{k+1}\equiv  Z^A_{2k}Z^A_{2k+1}Z^A_{2k+2}~,
    \label{eq:logical_operator_A}
\end{equation}
where $X^A_j/Z^A_j$ act on the $j$th red-colored qubit in Fig.~\ref{fig:cut}.
Here, $\tau^x_{k}$ creates a pair of domain walls at $k\pm\frac{1}{2}$, which is identified as two $m$-anyons, and  $\tau^z_{k}$ creates $\ZZ_2$ charge at $k$, identified as $e$-anyon.
Note that the relation  $\hat{1} = \prod_k \tau^z_k \tau^z_{k+1} = \prod_k Z^A_{2k+1}$ is not an operator identity --- rather, it holds within the ground-state subspace $\HH_{\partial A}$: $\prod_k Z^A_{2k+1} = \prod_{p \in A} \prod_{i \in p} Z_i$ measures the total $m$-anyon parity in $A$, which vanishes in $\HH_{\partial A}$.
Further, a closer analysis reveals that there are only $(L-1)$ independent zero-modes.
This reduction stems from the constraint where the number of $e$-anyon equals $0$ in $\HH_{\partial A}$:
\begin{equation}
    \prod_{k=1}^L\tau^x_{k}=\prod_{v\in A}\prod_{i\in v}X_i=\hat{1}~.
    \label{eq:partialA_e_z2}
\end{equation}
Therefore, $\HH_{\partial A}$ can be viewed as a 1D qubit chain of length $L$ subject to a $\ZZ_2$ global constraint in Eq.~\eqref{eq:partialA_e_z2}.

Similarly, the ground subspace $\mathcal{H}_{\partial \bar{A}}$ also describes a 1D qubit chain, with corresponding Pauli operators
\begin{equation}
        \bar{\tau}^{x}_{k}\equiv X_{2k-1}^{\bar{A}}X_{2k}^{\bar{A}}X_{2k+1}^{\bar{A}}~,\quad
        \bar{\tau}^{z}_{k}\bar{\tau}^{z}_{k+1}\equiv Z_{2k+1}^{\bar{A}}~,
        \label{eq:logical_operator_A_bar}
\end{equation}
where $X^{\bar{A}}_j$ and $Z^{\bar{A}}_j$ act on the $j$th blue-colored qubit in Fig.~\ref{fig:cut}.
The $L$ local qubits obey the global $\ZZ_2$ constraint $\prod_{k=1}^L \bar{\tau}^x_{k}=\hat{1}$, indicating zero $e$-anyon inside $\bar{A}$ within $\HH_{\bar{A}}$.
In addition, the identity $\hat{1}=\prod_{k=1}^L \bar{\tau}^z_{k}\bar{\tau}^z_{k+1}=\prod_k Z_{2k+1}^{\bar{A}}$ holds in $\HH_{\partial \bar{A}}$, indicating zero $m$-anyon.

\begin{figure}[!ht]
\centering
\includegraphics[width=0.2\textwidth]{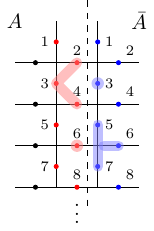}
\caption{
    Bipartition of the toric code model on a square lattice.
    Qubits at $\partial A~(\partial\bar{A})$ are denoted by red~(blue) dots.
    Regions $A$ and $\bar{A}$ are simply connected; the zoomed view shows qubit indices along the entanglement cut.
}
\label{fig:cut}
\end{figure}

The ground states of the full $H$ are now obtained by solving ground states of $H_{\partial A\cup\partial\bar{A}}$ within $\mathcal{H}_{\partial A}\otimes\mathcal{H}_{\partial \bar A}$.
In terms of $\tau_{A}$ and $\tau_{\bar{A}}$, $H_{\partial A\cup\partial\bar{A}}$ takes form
\begin{equation}
    H_{\partial A\cup\partial \bar A}=-\sum_{k=1}^{L}(\tau^{x}_{k}\bar{\tau}^{x}_{k}+\tau^{z}_{k}\tau^{z}_{k+1}\bar{\tau}^{z}_{k}\bar{\tau}^{z}_{k+1})~,
    \label{eq:cut_ham}
\end{equation}
subject to the global $\ZZ_2\times\ZZ_2$ constraint $\prod_{k}\tau^x_{k}=\prod_{k}\bar{\tau}^x_{k}=\hat{1}$.
Its ground state $\ket{\psi}$ is completely determined by minimizing all local terms in Eq.~\eqref{eq:cut_ham}, together with the $\ZZ_2\times\ZZ_2$ constraint.
Tracing out $\tau_{\bar A}$ yields RDM for region $A$, expressed in the $\tau$ basis.
This RDM, denoted $\rho_{\partial A}$, satisfies the local constraints
\begin{equation}
    [\tau_{k}^x,\rho_{\partial A}]=[\tau_{k}^z\tau_{k+1}^z,\rho_{\partial A}]=0~,
    \label{eq:toric_code_rdm_local_constraints}
\end{equation}
as well as \emph{a $\ZZ_2$ strong symmetry} due to the global constraint of $\HH_{\partial A}$:
\begin{equation}
\rho_{\partial A}=\prod_{k=1}^L\tau_{k}^x\cdot \rho_{\partial A} =\rho_{\partial A}\cdot \prod_{k=1}^L\tau_{k}^x~.
\label{eq:toric_code_rdm_strong_sym}
\end{equation}
These constraints lead to a unique solution:
\begin{equation}
    \rho_{\partial A}=\frac{1}{2^L} \left( \hat{1}+\prod_{k}\tau^x_{k} \right)
    \label{eq:rdm_toric_code}
\end{equation}

The above procedure to express RDM $\rho_A$ as a 1D mixed state $\rho_{\partial A}$ generalized readily to $\ZZ_2$ topological ordered systems beyond the exact solvable point.
We sketch the generalization here, and provide a more rigorous derivation using the framework of projected entangled pair states~(PEPS) in supplemental section I\cite{SM}.

The full system can be decomposed as $H_A+H_{\bar{A}}+H_{\partial A\cap\partial{\bar{A}}}$, where $H_{\partial A\cap\partial\bar{A}}$ contains all terms localized within a correlation length $\xi$ of the entanglement cut.
To construct the ground state, we first solve low-energy subspaces $\HH_{\partial A}$ and $\HH_{\partial\bar{A}}$ of regions $A$ and $\bar{A}$.
As in the exact solvable case, $\HH_{\partial A}$ supports anyonic excitations near the cut, and is effectively described by a 1D $\ZZ_2$ Ising chain.
Here, the $e$-anyon corresponds to a $\ZZ_2$ charge $\tau^z$, and the $m$-anyon corresponds to a $\ZZ_2$ domain wall --- an endpoint of a $\tau^x$-string. While we use Pauli notation $\tau^{x,z}$ for clarity, the essential structure for a generic model is that the effective boundary operators satisfy the Pauli algebra: $(\tau^z)^2 = (\tau^x)^2 = \hat{1}$ encodes the $\ZZ_2$ fusion rules of $e$ and $m$, while $\{\tau^x, \tau^z\} = 0$ reflects their semionic mutual statistics.

The absence of excitations in region $A$ imposes constraints on $\HH_{\partial A}$:
the $m$-loop and $e$-loop operators along the cut must act as identity in $\HH_{\partial A}$.
Expressed in terms of the effective $\ZZ_2$ Ising variables near the cut, the $m$-loop constraint imposes periodic boundary condition~(PBC), and the $e$-loop constraint restricts $\HH_{\partial A}$ within the $\ZZ_2$ even sector $\prod_j\tau^x_{j}=\hat{1}$.
An analogous description applies for $\HH_{\partial\bar{A}}$.

When restricted to $\HH_{\partial A}\otimes\HH_{\partial\bar{A}}$, the full ground state $\ket{\Psi}$ is approximated by the ground state $\ket{\psi}$ of $H_{\partial A\cap \partial\bar{A}}$.
Since the full system is in a uniform topological phase, anyons can traverse the entanglement cut.
This is realized by condensing bound pairs of $e$- and $m$-anyons that bridge the two sides of the cut, while leaving individual anyons deconfined.
Translating to the Ising chain language, this implies condensing $\tau^z\bar{\tau}^z$ and proliferating the $\tau^x\bar{\tau}^x$-string:
\begin{equation}
\lrangle{\tau_{i}^z\bar{\tau}_i^z\tau_{j}^z\bar{\tau}_j^z}\xrightarrow{\abs{i-j}\to\infty}C_e~,~
    \lrangle{\prod_{k=i}^j\tau_{k}^x\bar{\tau}_k^x}\xrightarrow{\abs{i-j}\to\infty} C_m
\label{eq:1D_psi_anyon_correlator}
\end{equation}
The long-range correlation of $\tau^z\bar{\tau}^z$~(creates $\ZZ_2\times\ZZ_2$ charge) signals that the $\ZZ_2\times\ZZ_2$ symmetry is spontaneously broken to some subgroup, while the non-decaying $\tau^x\bar{\tau}^x$-string~(creates diagonal $\ZZ_2$ domain walls) confirms that the diagonal $\ZZ_2$ subgroup remains unbroken.
Hence, $\ket{\psi}$ realizes a $\ZZ_2\times\ZZ_2\to\ZZ_2$ SSB phase.

Tracing out $\HH_{\partial\bar{A}}$ yields the 1D mixed state $\rho_{\partial A}$, which holds a $\ZZ_2$ strong symmetry $\prod_k\tau^x_k$.
The SSB pattern of $\ket{\psi}$ translates into the simultaneous exponential decay of both the diagonal $\ZZ_2$ charged-operator correlator and the off-diagonal $\ZZ_2$ domain walls:
\begin{equation}
    \Tr(\rho_{\partial A}\tau_{i}^z\tau_{j}^z)\sim\ee^{-\abs{i-j}/\xi_e}~,~
    \Tr(\rho_{\partial A}\prod_{k=i}^j\tau_{k}^x)\sim\ee^{-\abs{i-j}/\xi_m}
\label{eq:rdm_sw_ssb_anyon_correlator}
\end{equation}
In terms of anyons, Eq.~\eqref{eq:rdm_sw_ssb_anyon_correlator} is interpreted as neither $e$ nor $m$ is condensed, and $\xi_{e}$ and $\xi_{m}$ are the correlation lengths of $e$~($m$) anyons, respectively.
In the exact solvable limit, these correlators vanish strictly due to Eq.~\eqref{eq:toric_code_rdm_local_constraints}.

\emph{The 1D SW-SSB phase.}
In summary, the 1D RDM $\rho_{\partial A}$ of a generic $\ZZ_2$ topological ordered state exhibits the following universal properties:
\begin{enumerate}
    \item $\ZZ_2$ strong symmetry, indicating zero $e$;\label{enu:strong_sym}
    \item PBC, signaling zero $m$;\label{enu:pbc}
    \item Exponential decay of both $\ZZ_2$ charge and $\ZZ_2$ domain-wall correlators~(Eq.~\eqref{eq:rdm_sw_ssb_anyon_correlator}), signaling anyon deconfinement.\label{enu:exp_decay_charge_dw}
\end{enumerate}
Notably, property \ref{enu:exp_decay_charge_dw} cannot occur in any pure ground state(s) of a $\ZZ_2$ symmetric Ising chains\cite{levin2020constraints}.

We define the 1D strong $\ZZ_2$ symmetric $\rho_{\partial A}$ whose correlation function satisfies Eq.~\eqref{eq:rdm_sw_ssb_anyon_correlator} as the \emph{strong-to-weak spontaneous symmetry breaking}~(SW-SSB) mixed-state phase.
Physically, the exponential decay of $\ZZ_2$ domain wall operators signals that the strong $\ZZ_2$ symmetry is broken.
Meanwhile, operator $\tau^z$ carries charge under the induced $\ZZ_2$ weak symmetry (the diagonal subgroup of $\ZZ_2\times \ZZ_2$); hence, the exponential decay of $\lrangle{\tau^z_i\tau^z_j}$ indicates that the $\ZZ_2$ weak symmetry remains unbroken.
Our definition of SW-SSB --- based solely on \emph{linear} observables in $\rho_{\partial A}$ --- is distinct from characterizations via fidelity correlators~\cite{Lessa2025SWSSB} or Wightman functions~\cite{Liu2025diagnosingSWSSB,weinstein2024swssb} involving $\sqrt{\rho}$.
Such linear observables are directly accessible in experiments and numerical simulations, whereas extracting $\sqrt{\rho}$ in general requires full state tomography.
A detailed comparison between these approaches is provided in supplemental section II.

A remarkable consequence of this definition is that the 1D $\ZZ_2$ SW-SSB phase with PBC is self-dual under the Kramers-Wannier (KW) transformation.
Physically, KW duality exchanges the $\ZZ_2$ charge with the $\ZZ_2$ domain-wall:
\begin{equation}
    \tau^z_k\tau^z_{k+1}\to\tau^x_{k+1}~,\quad
    \tau^x_{k}\to\tau^z_{k}\tau^z_{k+1}~,
    \label{eq:kw_transf}
\end{equation}
For a mixed-state satisfying Eq.~\eqref{eq:rdm_sw_ssb_anyon_correlator}, the KW transformation simply interchanges $\xi_e$ and $\xi_m$, preserving the short‑range character of both correlators.
Globally, the KW transformation maps the even-$\ZZ_2$-charge condition $\prod_k\tau_k^x=1$~($\ZZ_2$ strong symmetry) to the even-domain-wall condition $\prod_{k}\tau^z_{k}\tau^z_{k+1}=1$~(PBC $\tau^z_{L+1}\equiv\tau_1^z$).
Consequently, a mixed state $\rho_{\partial A}$ in the 1D $\ZZ_2$ SW-SSB phase with PBC is self-dual.
Such self-duality has a natural bulk interpretation: it reflects the $e \leftrightarrow m$ electromagnetic duality of the toric code~(see details in supplemental section III).

The pure--mixed correspondence extends to all three phases of $\ZZ_2$ gauge theory~\cite{Fradkin1979}, completing the dictionary:
\begin{itemize}
    \item \textbf{Deconfined phase} $\longleftrightarrow$ \textbf{SW-SSB phase}: both the charged-operator and domain-wall correlators decay exponentially (Eq.~\eqref{eq:rdm_sw_ssb_anyon_correlator}).
    \item \textbf{Confined phase} ($m$ condenses) $\longleftrightarrow$ \textbf{symmetric phase}: domain walls proliferate while the charged correlator decays.
    \item \textbf{Higgs phase} ($e$ condenses) $\longleftrightarrow$ \textbf{SSB phase}: the $\ZZ_2$ charge condenses and long-range ordered while the domain wall decays; the strong symmetry is fully broken.
\end{itemize}
This three-way correspondence was first proposed in the PEPS context by Ref.~\cite{haegeman2015shadows}, where the 1D boundary MPS fixed points of the deformed toric code PEPS were discussed.
We provide a detailed derivation and numerical verification in supplemental sections~I and~IV.

\emph{Application: Detecting Symmetry Fractionalization}
The pure-mixed correspondence between 2D topological phases and 1D mixed-state phases is more than a formal mapping -- it provides a practical framework for extracting universal observables directly from a single ground state wavefunction.
Here, we generalize this framework to gapped $\ZZ_2$ spin liquids with global spin-rotation symmetry, where the $e$-anyon carries spin-$\frac{1}{2}$, identified as the fractionalized spinon.
The crucial new ingredient is the global spin-rotation symmetry, which manifests as a \emph{weak} symmetry of the 1D mixed state $\rho_{\partial A}$ — in contrast to the $\ZZ_2$ strong symmetry that encodes the topological order.
A classic example is the NN-RVB state~\cite{anderson1973} on a frustrated spin-$\frac{1}{2}$ system~\cite{anderson1973}.

While spinons are typically probed through the dynamic spin structure factor~\cite{Han2012} -- where they produce a low-energy continuum instead of sharp spin-wave peaks -- this spectral feature is often obscured by impurities experimentally~\cite{Savary2017}.
Our approach offers a ground-state alternative: we propose that the presence of gapped spinon spectrum can be detected through the spin rotation \emph{disorder parameter} $\abs{\lrangle{U_A(\theta)}}$ in a large subregion $A$, where
\begin{equation}
    \abs{\lrangle{U_A(\theta)}}\equiv \abs{\Tr[\rho_A U_A(\theta)]}~,~U_A(\theta)=\prod_{j\in A}\exp(\ii\theta S^z_j)
\end{equation}
In symmetric phases, it obeys an area law~\cite{Kadanoff1971disorder}, where $\abs{\lrangle{U_A(\theta)}}\sim\ee^{-f(\theta) L}$, for some function $f(\theta)$.
Crucially, when the $e$-anyon carries half-integer spin, $f(\theta)$ develops \emph{non-smooth behavior} as a function of $\theta$.
Numerical evidence supporting this prediction, obtained from NN-RVB state on the kagome lattice, is presented in Fig.~\ref{fig:disorder_param_rvb}~(details are given in supplemental section IV).

\begin{figure}[t]
    \centering
    \includegraphics[scale=0.57]{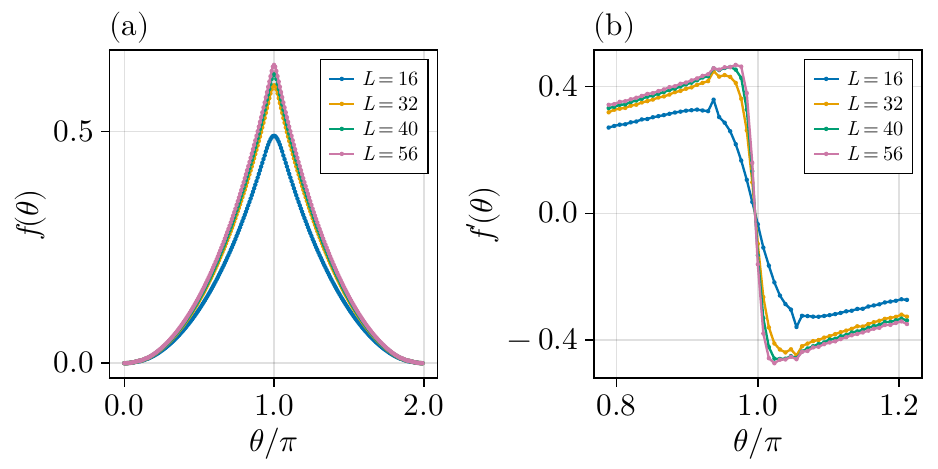}
    \caption{Disorder parameters of the NN-RVB state on the kagome lattice. The non-smoothness at $\theta=\pi$ corresponds to a discontinuity in the derivative $f'(\theta)$.}
    \label{fig:disorder_param_rvb}
\end{figure}

The above result can be derived analytically using the matrix product density operator (MPDO) formalism.
The logic of the derivation is as follows.
We encode $\rho_{\partial A}$ as an MPDO and impose the $\ZZ_2$ SW-SSB and $U(1)$ symmetry conditions as constraints on the local tensor $M$.
The disorder parameter $\lrangle{U_A(\theta)}$, when evaluated in the thermodynamic limit, equals the dominant eigenvalue of a $\theta$-twisted transfer matrix $T(\theta)$.
The symmetry fractionalization condition forces an eigenvalue crossing between $\theta = 0$ and $\theta = 2\pi$, and the Hermiticity of $\rho_{\partial A}$ further pins the crossing to $\theta = \pi$.
The detailed derivation follows.

As in the case of pure $\ZZ_2$ topological order, $\rho_{\partial A}$ of a gapped $\ZZ_2$ spin liquid also belongs to the $\ZZ_2$ SW-SSB phase: it possesses a strong $\ZZ_2$ symmetry generated by the $m$-anyon loop, denoted as $J_{\partial A}$~(Eq.~\eqref{eq:toric_code_rdm_strong_sym}), and exhibits short-range correlation condition in Eq.~\eqref{eq:rdm_sw_ssb_anyon_correlator}.
The key new ingredient for spin liquids is the incorporation of global spin rotation symmetry $U(\theta)=\ee^{\ii S^z\theta}$.
This global symmetry manifests as a weak $U(1)$ symmetry on RDM $\rho_A$.
Denoting its representation in the entanglement space as $W_{\partial A}(\theta)=\prod_{j\in A}W_j(\theta)$, we have $[\rho_{\partial A},W_{\partial A}(\theta)]=0$.
For the NN-RVB case, the $e$-anyon carries half-integer spin, identified as the spinon.
Therefore, $W_{\partial A}(2\pi)\circ e=-e$, identifying it as the $m$-anyon loop:
\begin{equation}
    W_{\partial A}(2\pi)=J_{\partial A}
    \label{eq:set_spinon}
\end{equation}

With these properties in place, we make a further simplifying assumption: the 1D mixed state $\rho_{\partial A}$ is well approximated by a matrix‑product density operator (MPDO):
\begin{equation}
    \rho_{\partial A}=\Tr[M^{L_{\partial A}}]=
    \adjincludegraphics[scale=1,valign=c]{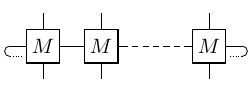}
    \label{}
\end{equation}
where $M$ denotes the local tensor of the MPDO.
Hence, the disorder parameter reads
\begin{equation}
\begin{aligned}
    \lrangle{U_A(\theta)}&=\Tr\big(\rho_{\partial A}\cdot W_{\partial A}(\theta)\big)
            = \Tr\big(T(\theta)^{L_{\partial A}}\big)\\
    &=
    \adjincludegraphics[scale=1,valign=c]{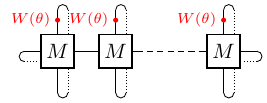}
\end{aligned}
\end{equation}
where the $\theta$-twisted transfer matrix is $ T(\theta)=
\adjincludegraphics[scale=1,valign=c]{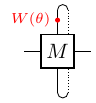}
    $.
In the thermodynamic limit $L_{\partial A}\to\infty$, we have
\begin{equation}
f(\theta)=\log\abs{U_A(\theta)}=-\log\abs{\alpha_{\rm max}(\theta)}
\label{eq:disorder_param_transfer_mat_eigs}
\end{equation}
where $\alpha_{\rm max}(\theta)$ is the eigenvalue of $T(\theta)$ with the largest modulus.

To proceed, we implement the symmetry of $\rho_{\partial A}$ on the MPDO, which manifests as gauge transformation on virtual legs of local tensor $M$.
First, the $\ZZ_2$ strong symmetry $\rho_{\partial A}=J_{\partial A}\cdot\rho=\rho\cdot J_{\partial A}$ leads to
\begin{equation}
    \adjincludegraphics[scale=1,valign=c]{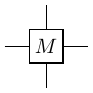}
    =
    \adjincludegraphics[scale=1,valign=c]{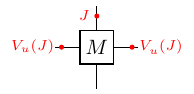}
    =
    \adjincludegraphics[scale=1,valign=c]{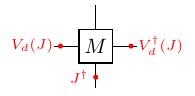} .
    \label{eq:teneq_z2_strong_sym}
\end{equation}
Physically, $V_u(J)$ ($V_d(J)$) creates an $m$-anyon in the ket (bra) layer.
To capture the $\ZZ_2$ SW-SSB for $\rho_{\partial A}$ in MPDO, a virtual leg $\ZZ_2$ symmetry is introduced (see supplemental section V for more details)
\begin{equation}
\begin{aligned}
\adjincludegraphics[scale=1,valign=c]{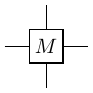}
   =
     \adjincludegraphics[scale=1,valign=c]{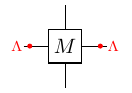} \\,
\end{aligned}
    \label{eq:teneq_igg}
\end{equation}
where $\Lambda^2 = \hat{1}$, and
\begin{equation}
\Lambda\cdot V_{u/d}(J)=- V_{u/d}(J)\cdot \Lambda
\label{eq:lambda_vj_anticomm}
\end{equation}
Physically, $\Lambda$ creates $e$-anyon at both bra and ket space, and its anti-commutation with $V_{u/d}(J)$ in Eq.~\eqref{eq:lambda_vj_anticomm} reflects the mutual semionic statistics between $e$ and $m$.
As shown in supplemental section V, imposing Eqs.~\eqref{eq:teneq_z2_strong_sym} and \eqref{eq:teneq_igg} immediately imply the exponential decay of both charge and domain walls.

Now, imposing the global $U(1)$ spin rotation symmetry, which further constrains $M$ as
\begin{equation}
    \adjincludegraphics[scale=1,valign=c]{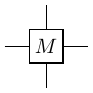}
    =
    \adjincludegraphics[scale=1,valign=c]{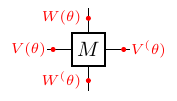} .
    \label{eq:teneq_u1_weak_sym}
\end{equation}
The algebraic relations for gauge transformation $V$'s follow those on physical legs of the MPDO.
In particular, $V(2\pi)=V_u(J)V_d(J)$ due to Eq.~\eqref{eq:set_spinon}.

Symmetries of $\theta$-twisted transfer matrix can be derived from the above equations as
\begin{equation}
    T(\theta)=\Lambda\cdot T(\theta)\cdot\Lambda^\dg
    =V(\theta')\cdot T(\theta)\cdot V^\dg(\theta')
    \label{eq:twisted_tm_sym}
\end{equation}
for any angle $\theta'$.

Equipped with these symmetry properties, we now investigate the flow of the dominant eigenvalue for $T(\theta)$, which determines the behavior of disorder parameter $f(\theta)$.
When $\theta=0$, the transfer matrix $T(0)$ holds non-degenerate dominant eigenvalue.
Let $\vket{r_1(0)}$ be the leading eigenvector of $T(0)$ with eigenvalue $\alpha_{0}(0)$.
We track the smooth trajectory of $\alpha_0(\theta)$ starting from $\theta=0$ to $\theta=2\pi$
\begin{equation}
T(\theta)\vket{r_0(\theta)}=\alpha_{0}(\theta)\vket{r_0(\theta)}.
\label{eq:alpha_theta_flow}
\end{equation}
where $\alpha_0(\theta)$ and $\ket{r_0(\theta)}$ are smooth.

Due to the fractionalized spinon feature where $W(2\pi)=J$ (Eq.~\eqref{eq:set_spinon}), $T(\theta+2\pi)$ are then related to $T(\theta)$ by a gauge transformation:
\begin{equation}
    T(\theta+2\pi)=V_{u}^\dg(J)\cdot T(\theta)\cdot V_{u}(J)
    \label{eq:twisted_tm_2pi_gauge_transf}
\end{equation}
Therefore, $V_u^\dg(J)\ket{r_0(0)}$ is then non-degenerate dominant eigenvector of $T(2\pi)$, where
\begin{equation}
   T(2\pi)\cdot V_u^\dg(J)\ket{r_0(0)}=\alpha_0(0)\cdot V_u^\dg(J)\ket{r_0(0)}
   \label{}
\end{equation}
Comparing with Eq.~\eqref{eq:alpha_theta_flow}, one might identify $V_u^\dg(J)\ket{r_0(0)}$ with $\ket{r_0(2\pi)}$.
However, we will show that these two states are distinct.
The continuity of $\ket{r_0(\theta)}$ implies that $\ket{r_0(2\pi)}$ and $\ket{r_0(0)}$ share the same quantum number under symmetry $\Lambda$.
However, due to the anticommutation relation in Eq.~\eqref{eq:lambda_vj_anticomm}, $V_u^\dg(J)\ket{r_0(0)}$ hosts opposite quantum number under $\Lambda$ from $\ket{r_0(0)}$.
Therefore, we conclude that $\abs{\alpha_0(2\pi)}<\abs{\alpha_{\max}(2\pi)}=\abs{\alpha_0(0)}$, and the exchange of the dominant eigenvalue in general leads to the non-smooth $f(\theta)$ due to Eq.~\eqref{eq:disorder_param_transfer_mat_eigs}, consistent with Fig.~\ref{fig:disorder_param_rvb}.

We note that there is another key feature in Fig.~\ref{fig:disorder_param_rvb}(a), where the non-smooth point is at $\theta=\pi$.
In terms of $T(\theta)$, such behavior means that the largest two eigenvalues of $T(2\pi)$ crosses at $\theta=\pi$, and is degenerate.
To analytically derive this, we further consider the Hermiticity condition $\rho=\rho^\dg$, which imposes local tensor constraints:
\begin{equation}
    \adjincludegraphics[scale=1,valign=c]{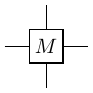}
    =
    \adjincludegraphics[scale=1,valign=c]{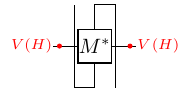}
    \label{eq:teneq_hermiticity}
\end{equation}
where $[V(H)\KK]^2 = \hat{1}$ and $V(H)\KK \cdot V_u(J) = V_d(J) \cdot V(H)\KK$, with $\KK$ complex conjugation.
Hence,
\begin{equation}
    T(\theta)=V(H)\cdot T^*(-\theta)\cdot V^\dg(H)
\end{equation}
Together with Eq.~\eqref{eq:twisted_tm_2pi_gauge_transf}, we conclude
\begin{equation}
    T(\pi)=V(H)\KK V_u(J)\cdot T(\pi) \cdot [V(H)\KK V_u(J)]^\dg
    \label{}
\end{equation}
From Eq.~\eqref{eq:lambda_vj_anticomm}, $\{V(H)\KK V_u(J),\Lambda\}=0$.
Hence, $\ket{\alpha_0(\pi)}$ and $V(H)\KK V_u(J)\ket{\alpha_0(\pi)}$ are distinct degenerate eigenstates of $T(\pi)$, as they have different quantum numbers under $\Lambda$.
This explains the non-smoothness at $\theta=\pi$ in Fig.~\ref{fig:disorder_param_rvb}(a).

As a further application of this framework, the MPDO analysis can be extended to other SET phases.
In supplemental section~VI, we show that for a $\ZZ_2$ SET where the $e$-anyon carries a fractional $\ZZ_2$ quantum number, the $m$-anyon correlator in the presence of a $\ZZ_2$ disorder operator becomes long-range ordered:
\begin{equation}
    \frac{\lrangle{m_i m_j U_A(g)}}{\lrangle{U_A(g)}}\xrightarrow{\abs{i-j}\to\infty} O(1)
\label{}
\end{equation}
with $i$ and $j$ located near $\partial A$.

\emph{Summary and discussion.}
The central message of this work is that the RDM of a 2D topological phase, evaluated at an entanglement cut, realizes a 1D nontrivial mixed-state phase, whose symmetry and correlation structure directly encodes universal topological data.
For the $\ZZ_2$ toric code, the RDM at the entanglement cut realizes a 1D $\ZZ_2$ SW-SSB phase, whose self-duality under the Kramers--Wannier transformation reflects the electromagnetic duality of the bulk topological order.
By extending this framework to the gapped $\ZZ_2$ spin liquids, we propose a direct ground-state diagnostic of spin-rotation symmetry fractionalization, where the corresponding disorder parameter develops non-analytic cusp at $\theta=\pi$.
Analytical derivations based on MPDO and numerical results for the kagome-lattice RVB state are provided to corroborate this prediction.

Looking forward, our work suggests several promising directions.
The linear correlators definition of SW-SSB phase readily generalizes to $d$-dimensional mixed states with a strong $n$-form symmetry\cite{zhang2025strong}.
In this case, both the order parameter $\lrangle{O_c}$ (supported on a closed $n$-dimensional manifold $M_c$) and the disorder parameter $\lrangle{O_d}$ (supported on a $(d-n)$-dimensional manifold with boundary $M_d$) obey ``volume-law'' scaling:
\begin{equation}
  -\ln\abs{\lrangle{O_c}}\propto V^c_{n+1}~,~
  -\ln\abs{\lrangle{O_d}}\propto V^d_{d-n}~,
\end{equation}
where $V^c_{n+1}$ is the minimal volume of manifolds bounded by $M_c$, and $V^d_{d-n}$ is the volume of $M_d$.

The natural extension of the pure--mixed correspondence to topological phases with non-invertible~(categorical) symmetries~\cite{JiWen2020,kong2020algebraic} is particularly compelling, as the framework may shed new light on detection of such exotic orders.
Extending these ideas to chiral topological phases remains an important open problem.

The most immediate implication of this work is experimental: the disorder parameter $\lrangle{U_A(\theta)}$ is a concrete, measurable observable that can be accessed on quantum simulation platforms.
It requires only a single ground-state wavefunction, making it a practical route to detecting signatures of fractionalized excitations in engineered quantum systems.

\emph{Acknowledgment.}
We thank Pengfei Zhang, Yingfei Gu, Meng Cheng, and Yunqin Zheng for helpful discussions.
This work is supported by MOST NO. 2022YFA1403902, NSFC Grant No.12574179.

\bibliography{bibpaper}

\clearpage
\setcounter{equation}{0}
\renewcommand{\theequation}{S\arabic{equation}}

\setcounter{table}{0}
\renewcommand{\thetable}{S\arabic{table}}
\renewcommand{\theHtable}{\thetable}

\setcounter{figure}{0}
\renewcommand{\thefigure}{S\arabic{figure}}
\renewcommand{\theHfigure}{\thefigure}

\setcounter{section}{0}
\setcounter{secnumdepth}{3}

\title{Supplemental Materials: From Topological Order to Mixed-State Phases: \\A Ground-State Probe of Fractionalized Excitations}

\maketitle

\onecolumngrid
\tableofcontents
In this supplemental material, the RDM of the toric code phase beyond the fixed-point wavefunction is discussed in Sec.~\ref{app:rdm_beyond_fp}.
Sec.~\ref{app:diff_def_sw_ssb} discusses the connection and difference between the SW-SSB phases we proposed here, and the definition proposed previously.
Sec.~\ref{app:dual_peps_kw} derive the Kramers-Wannier self-duality for the 1D SW-SSB phase from the bulk perspective.
Sec.~\ref{app:peps_rvb_numerics} gives the numerical details for anyon correlators in deformed toric code state and disorder parameter in the kagome RVB state.
Sec.~\ref{app:mps_mpdo_correlator} relates the virtual leg symmetries to the behavior of charge and domain wall correlators in MPDO.
Finally, Sec.~\ref{app:z2_set_imo} discusses universal feature for $\ZZ_2$ symmetric enriched topological ordered phases.

\section{RDM for the toric code phase beyond the fixed point}\label{app:rdm_beyond_fp}

In the main text, we established the pure--mixed correspondence at the exactly solvable (fixed-point) limit of the toric code.
In this section, we extend the analysis beyond the fixed point in two complementary ways.
First, we show that the correspondence is stable under finite-depth local unitaries (FDLU).
Second, we derive the same correspondence from the projected-entangled-pair state (PEPS) framework, where the entanglement space is identified with the virtual legs at the entanglement cut.

\subsection{Finite-depth local unitary}
Here, we show that the pure--mixed correspondence is stable against FDLU~(denoted as $V$).
Let $V$ consist of $d$ layers of local unitaries, each with linear size up to $n$.
Under conjugation by $V$, each term in the transformed Hamiltonian $H'$ is supported on a region of width $l\sim 2nd$.

\begin{figure}[!ht]
\centering
\adjincludegraphics[scale=1,valign=c]{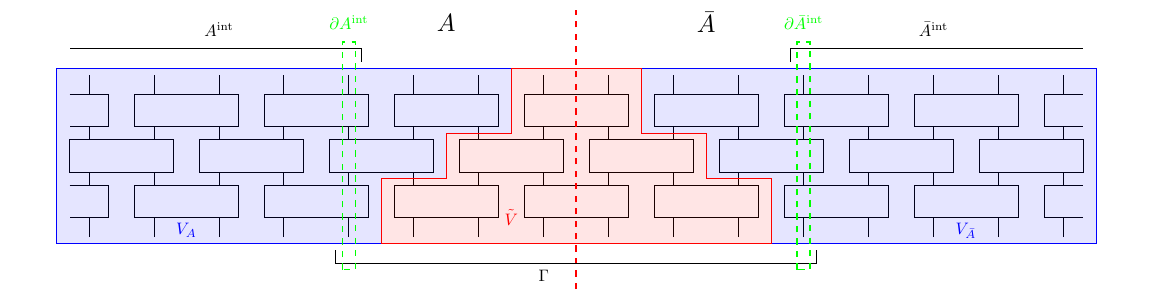}
\caption{Decomposition of the FDLU $V$. The time direction runs from bottom to top.
    Red dashed line: entanglement cut. 
    Region left (right) to the cut: $A$ ($\bar A$). 
    Regions in three square brackets: $A^{\mathrm{int}}$, $\bar A^{\mathrm{int}}$ and $\Gamma$, supporting $H_{A^{\mathrm{int}}}$, $H_{\bar A^{\mathrm{int}}}$ and $H_{\Gamma}$ respectively.
    Green dashed rectangles: $\partial A^{\mathrm{int}}=\Gamma\cap A^{\mathrm{int}}$ and $\partial \bar A^{\mathrm{int}}=\Gamma\cap\bar A^{\mathrm{int}}$.
    Blue: $V_A$ and $V_{\bar A}$ (gates acting entirely within $A^{\mathrm{int}}$ and $\bar A^{\mathrm{int}}$).
    Red: $\tilde V$ (gates coupling across the cut).
}
\label{fig:FDLU_3parts}
\end{figure}

We first define the spatial regions (see Fig.~\ref{fig:FDLU_3parts}).
The FDLU $V$ mixes degrees of freedom within a distance $\sim nd$ of the entanglement cut.
Let $A^{\mathrm{int}} \subset A$ be the set of sites in $A$ whose distance from the cut exceeds $nd$, so that gates in $V$ acting within $A^{\mathrm{int}}$ never cross into $\bar A$.
Define $\bar A^{\mathrm{int}} \subset \bar A$ similarly.
$\Gamma$ is a strip of width $\sim 2nd$ straddling the cut, such that a term $\prod_{i\in p}Z_i$ ($\prod_{i\in v}X_i$) in the exactly solvable Hamiltonian is sent across the cut by the FDLU $V$ iff $p\subseteq\Gamma$ ($v\subseteq\Gamma$).
The boundary of $A^{\mathrm{int}}$ adjacent to $\Gamma$ is denoted $\partial A^{\mathrm{int}}=\Gamma\cap A^{\mathrm{int}}$, and analogously $\partial \bar A^{\mathrm{int}}$ for $\bar A^{\mathrm{int}}$.
By construction, $\partial A^{\mathrm{int}}$ is a narrow 1D stripe, where qubits within it cannot form complete local terms of the original Hamiltonian --- they only appear as part of a local term that straddles $\Gamma$.

With these regions defined, we decompose the original toric code Hamiltonian $H$ into three commuting parts (see Fig.~\ref{fig:FDLU_3parts}).
$H_{A^{\mathrm{int}}}$ ($H_{\bar A^{\mathrm{int}}}$) collects all stabilizer terms whose support after conjugation by $V$ lies entirely within $A$ ($\bar A$); these terms are supported within $A^{\mathrm{int}}$ ($\bar A^{\mathrm{int}}$).
The remaining terms, supported within $\Gamma$, constitute $H_\Gamma$.
Correspondingly, the FDLU factorizes as $V = (V_A \otimes V_{\bar A}) \tilde V$, where $V_A$ ($V_{\bar A}$) is the subset of gates acting solely within $A^{\mathrm{int}}$ ($\bar A^{\mathrm{int}}$), and $\tilde V$ collects the remaining gates. 
The transformed Hamiltonian then reads
\begin{align} 
    H'&=VHV^{\dagger}=-\sum_{p}V\prod_{i\in p}Z_iV^{\dagger}-\sum_{v}V\prod_{i\in v}X_iV^{\dagger}\nonumber\\
    &\equiv H'_A+H'_{\bar A}+H'_{\Gamma}\nonumber\\
    &\equiv V_AH_{A^{\mathrm{int}}}V_A^{\dagger}+V_{\bar A}H_{\bar A^{\mathrm{int}}}V_{\bar A}^{\dagger}+VH_{\Gamma}V^{\dagger}
    \label{eq:toric_fdlu}
\end{align}
whose ground state is $\ket{\psi'}\equiv V|\psi\rangle = (V_A \otimes V_{\bar A})\tilde V |\psi\rangle$, with $\ket{\psi}$ the ground state of $H$.
Hence,
\begin{equation}
    \rho_A =\Tr_{\bar A}(|\psi'\rangle\langle\psi'|)=V_A\cdot \Tr_{\bar A}(\tilde V|\psi\rangle\langle\psi|\tilde V^{\dagger})\cdot V_A^{\dagger}.
\end{equation}
In the following, we will show that $\rho_A$ is isomorphic to a quasi-1D mixed-state.

Since the three terms in Eq.~\eqref{eq:toric_fdlu} commute, we can follow a similar strategy as in the main text.
The entanglement cut divides the system into two parts $\HH_A \otimes \HH_{\bar A}$.
We first consider the ground-state manifold of $H_{A^{\mathrm{int}}}$, as a subspace of $\HH_A$.
It has a natural quasi-1D (width $\sim nd$) realization, denoted as $\HH_{\partial A}$, which consists of the qubits in the strip $\Gamma \cap A$ (the region between $\partial A^{\mathrm{int}}$ and the entanglement cut), together with effective qubits described by $\tau$ operators on $\partial A^{\mathrm{int}}$.
To work out the $\tau$ spins, we note that $\partial A^{\mathrm{int}}$ contains exactly one closed $e$-loop ($C_e$) and one closed $m$-loop ($C_m$), giving the global constraints $\prod_{i\in C_e} Z_i = \prod_{i\in C_m} X_i = \hat{1}$.
Therefore, the effective $\tau$ spins are the same as those defined in Eqs.~(2) and~(3) of the main text.
This quasi-1D realization induces an isometry map $P_A : \HH_A \to \HH_{\partial A}$.
Further, since $H_A'$ and $H_{A^{\mathrm{int}}}$ are related by the FDLU $V_A$, the ground states of $H_A'$ can also be mapped to states in $\HH_{\partial A}$, with isometry $P_A V_A^\dagger$.

Similarly, the ground-state manifold $H_{\bar A}'$ are supported on a quasi-1D system $\HH_{\partial \bar{A}}$, with the isometry map $P_{\bar{A}}V_{\bar{A}}^\dg$ from $\HH_{\bar{A}}$.
$\HH_{\partial \bar{A}}$ contains effective $\bar{\tau}$ spins in $\partial\bar{A}^{\mathrm{int}}$, as well as all qubits in $\bar{A}\setminus \bar{A}^{\mathrm{int}}$.

Equipped with the above isometries, we can express $H_\Gamma'$ as a Hamiltonian in the coupled quasi-1D system $\mathcal{H}_{\partial A}\otimes\mathcal{H}_{\partial\bar A}$:
\begin{align}
    \tilde H_{\Gamma}&\equiv(P_A V_A^{\dagger})\otimes (P_{\bar A}V_{\bar A}^{\dagger})\cdot H'_{\Gamma}\cdot (V_AP_A^\dg)\otimes (V_{\bar A}P_{\bar A}^\dg)\nonumber\\
    &=(P_A\otimes P_{\bar A})\cdot \tilde V\cdot H_\Gamma\cdot\tilde V^{\dagger}\cdot (P_A^\dg\otimes P_{\bar A}^\dg)~,
\end{align}
where we use $\tilde V = (V_A^{\dagger}\otimes V_{\bar A} ^{\dagger})\cdot V$ in the second line.

The ground state of $\tilde H_{\Gamma}$ reads 
\begin{equation}
    \ket{\psi_{\Gamma}}=(P_AV_A^{\dagger})\otimes (P_{\bar A}V_{\bar A}^{\dagger}) \ket{\psi'} 
    = (P_A\otimes P_{\bar A})\cdot\tilde V\ket{\psi}.
\end{equation}
Therefore, we obtain the quasi-1D RDM $\rho_{\partial A}$
\begin{equation}
    \rho_{\partial A}\equiv \Tr_{\partial\bar{A}}(\ket{\psi_{\Gamma}}\bra{\psi_{\Gamma}})
    =P_A\cdot  \Tr_{\partial\bar A}(P_{\bar A}\tilde V|\psi\rangle\langle\psi|\tilde V^{\dagger}P_{\bar A}^\dg)\cdot P_A^\dg
\end{equation}

Note that in $\HH_{\partial A}$, $P_A^{\dagger}\prod_{i\in C_m} X_iP_A=\prod_{j}\tau^x_j=\hat 1$, indicating the number of $e$-string passing through $\partial A^{\mathrm{int}}$ must be even.
It generates strong $\ZZ_2$ symmetry of $\rho_{\partial A}$: $\rho_{\partial A} =  \prod_{j}\tau^x_j\cdot\rho_{\partial A} = \rho_{\partial A}\cdot\prod_{j}\tau^x_j$.

As a final step, we discuss the behaviour of charge and domain-wall correlators.
Here, $\tau^z$ carries odd $\ZZ_2$ charge, serving as the order parameter. 
Domain walls are identified as end points of $\tau^x$ strings.
Translating to the operators in $\mathcal{H}_{A}$, they are
\begin{align}
    \tau^{z}_{i}\tau^{z}_{j} \equiv P_A^{\dagger}\prod_{k\in C_{e,ij}} Z_kP_A,\ \ \  \prod_{k=i}^{j} \tau^{x}_{k} \equiv P_A^{\dagger}\prod_{k\in C_{m,ij}} X_k P_A,
\end{align}
where $C_{e, ij}$ is the $e$ string in $\partial A^{\mathrm{int}}$ connecting sites $i$ and $j$, and $C_{m, ij}$ is an $m$-string in $\partial A^{\mathrm{int}}$ connecting $i$ and $j$.
As the FDLU does not close the gap, $H'$ remains in the deconfined phase, where correlators of both $e$ and $m$ anyons decay exponentially.
More precisely, they vanishes beyond $r_{e/m}$, where $r_e$ and $r_m$ are the finite length scales induced by the FDLU.
In terms of $\tau$ spins, we then have
\begin{align}
    \Tr(\rho_{\partial A}\tau^{z}_{i}\tau^{z}_{j}) = \lrangle{e_{\partial A,i}\, e_{\partial A,j}} \sim f_e(\abs{i-j})\theta(-|i-j|+r_e),\\
    \Tr\Bigl(\rho_{\partial A}\prod_{k=i}^{j}\tau^{x}_{k}\Bigr) = \lrangle{m_{\partial A,i}\, m_{\partial A,j}} \sim f_m(\abs{i-j})\theta(-|i-j|+r_m),
\end{align}

\subsection{RDM of toric code phase from the PEPS framework}
We now present a complementary derivation of the pure/mixed correspondence using the PEPS formalism.
In PEPS, the entanglement degrees of freedom are identified as virtual legs crossing the entanglement cut, providing a natural one-dimensional structure.

\subsubsection{RDM in the entanglement space}\label{app:RDM-entanglement-space}
We start by considering a generic bipartite tensor wavefunction, represented diagrammatically as
\begin{equation}
        \ket{\psi}=\Tr_{\HH_{\partial A}} T_A\otimes T_{\bar A} = \adjincludegraphics[scale=1,valign=c]{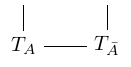},
\end{equation}
where the trace is performed over the entanglement space $\HH_{\partial A}$. 
For a 2D gapped state described by a PEPS, $\HH_{\partial A}$ comprises the virtual legs along the entanglement cut, where $\dim \HH_{\partial A} \sim \chi^{L_{\partial A}}$, with $\chi$ the virtual bond dimension.

The RDM $\rho_A$ supported in the physical Hilbert space $\HH_A$ can be mapped to a 1D mixed state $\rho_{\partial A}$ defined on $\HH_{\partial A}$~\cite{Cirac2011boundary,Schuch2013transfer}.
To obtain $\rho_{\partial A}$, we first define the environment tensors $\sigma_{A/\bar{A}}$ by contracting physical legs:
\begin{equation}
    \sigma_A = \Tr_{\HH_A} T_A\otimes T_A^{*}
    =
    \adjincludegraphics[scale=1,valign=c]{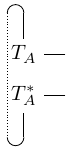}
    ~~ ,\quad  
    \adjincludegraphics[scale=1,valign=c]{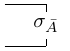}=
    \adjincludegraphics[scale=1,valign=c]{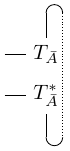}
\end{equation}
We then have $T_{\alpha} = S_{\alpha} \sqrt{\sigma_{\alpha}}$ through polar decompisition, with $S_{\alpha}$ isometries.
Therefore,
\begin{equation}
    \rho_A = \Tr_{\HH_{\bar A}}\ket{\psi}\bra{\psi}=
    \adjincludegraphics[scale=1,valign=c]{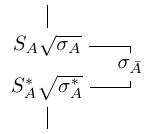}\Longrightarrow\rho_A =S_A\rho_{\partial A}S_A^\dagger ,
\end{equation}
where $\rho_{\partial A}$ is a 1D mixed state in $\HH_{\partial A}$, defined as
\begin{equation}
    \rho_{\partial A} = \sqrt{\sigma_A^T}\sigma_{\bar A}\sqrt{\sigma_A^T}=
    \adjincludegraphics[scale=1,valign=c]{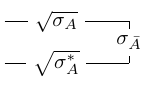}.
    \label{eq:rho_entanglement_space}
\end{equation}

\subsubsection{Fixed-point PEPS representation}\label{app:Fixed_pt_PEPS}
In this part, we will solve $\rho_{\partial A}$ for the toric code fixed-point wavefunction by its PEPS representaion.
Consider toric code model on the square lattice, 
\begin{equation}
    H_{\textrm{T.C.}} = -\sum_v \adjincludegraphics[scale=1,valign=c]{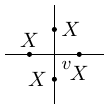} 
  -\sum_p  \adjincludegraphics[scale=1,valign=c]{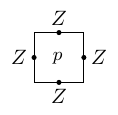}
  \label{eq:tc_ham}
\end{equation}
On the infinite plane, its ground state can be represented as a PEPS constructed from two types of local tensors:
\begin{equation}
     \adjincludegraphics[scale=1,valign=c]{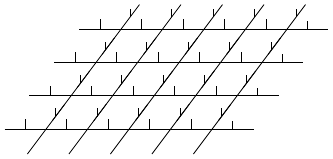} \quad V=\adjincludegraphics[scale=1,valign=c]{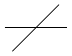}\quad \quad\quad E=\adjincludegraphics[scale=1,valign=c]{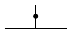}
\label{eq:TC_PEPS_tensor}
\end{equation}
Here, the site tensor $V$ possesses four virtual qubits, while the bond tensor $E$ has two virtual qubits and one physical qubit.
These tensors are uniquely defined (up to normalization) by the following local algebraic constraints:
\begin{equation}
\begin{aligned}
    V & = \adjincludegraphics[scale=1,valign=c]{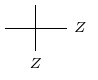}
= \adjincludegraphics[scale=1,valign=c]{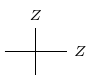}
= \adjincludegraphics[scale=1,valign=c]{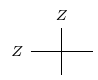}
=\adjincludegraphics[scale=1,valign=c]{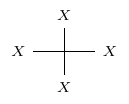}\\
E & =\adjincludegraphics[scale=1,valign=c]{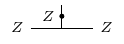}
= \adjincludegraphics[scale=1,valign=c]{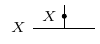}
=\adjincludegraphics[scale=1,valign=c]{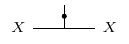}
\end{aligned}
\label{eq:toric_code_local_tensor_constraint}
\end{equation}
It is straightforward to verify that the physical wavefunction, obtained by contracting all virtual bonds, stabilizes every term in $H_{{\rm T.C.}}$. 

We now compute  $\rho_{\partial A}$ for a simply connected subregion $A$ using Eq.~\eqref{eq:rho_entanglement_space}.
$T_A$ is the restriction of the PEPS to the finite patch $A$, and $\HH_{\partial A}$ are the uncontracted dangling virtual legs.
The environment tensor $\sigma_A$ reads
\begin{equation}
    \sigma_A= 
    \adjincludegraphics[scale=1,valign=c]{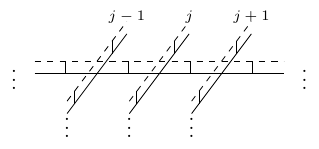}
\label{eq:sigma_A_tn}
\end{equation}
where dashed and solid lines denote the bra and ket layers of the contracted PEPS, respectively.
$\sigma_A$ satisfies constraints inherited from Eq.~\eqref{eq:toric_code_local_tensor_constraint}, where
\begin{equation}
    \sigma_A = Z_j Z_{j+1}\cdot \sigma_A\cdot Z_j Z_{j+1}
    = X_j\cdot \sigma_A\cdot X_j
    = \left(\prod_{j=1}^L X_j\right)\cdot\sigma_A,
    \label{eq:sigma_sol}
\end{equation}
which enforces a unique solution (up to normalization): $\sigma_A \propto \hat{1} + \prod_j X_j$.
Following similar procedure for $\bar{A}$, we have $\sigma_{\bar{A}}\propto\hat{1}+\prod_jX_j$.
Inserting these into Eq.~\eqref{eq:rho_entanglement_space} yields
\begin{equation}
    \rho_{\partial A}\propto \hat{1} + \prod_j X_j,
\end{equation}
which in perfect agreement with the result in main text.
Crucially, $\rho_{\partial A}$ exhibits a strong  $\ZZ_2$ symmetry:
\begin{equation}
    \rho_{\partial A} = \prod_{j} X_j\cdot \rho_{\partial A} = \rho_{\partial A} \cdot \prod_{j} X_j.
    \label{eq:fp_strongz2}
\end{equation}

\emph{Anyon string operators in PEPS.---}
We now examine the behavior of anyonic excitations.
The toric code realizes the deconfined phase of a $\ZZ_2$ gauge theory, whose low-energy excitations include gauge charges ($e$), fluxes ($m$), as well as their bound states.
In the exact solvable point, they are created by string operators: a $Z$-string along a direct lattice path $\gamma$, $(W_e)_\gamma =\prod_{j\in\gamma}Z_j$, creates two $e$-particles at its endpoints, whereas an $X$-string along a dual lattice path $\gamma^*$, $(W_m)_{\gamma^*}=\prod_{j\in \gamma^{*}}X_j$, creates two $m$-particles. 

Remarkably, by utilizing the tensor constraints in Eq.~\eqref{eq:toric_code_local_tensor_constraint}, $W_m$ is deformed into an $X$-string purely on the virtual legs of the PEPS, which depicted by the red dashed line in the figure below.
Similarly, $W_e$ reduces to two localized $Z$ operators on the virtual legs, marked by blue dots.
The mutual $\pi$-statistics between the $e$ and $m$ anyons is captured by the anti-commutation relation $\{X,Z\}=0$ on the virtual legs.
\begin{equation}
     \adjincludegraphics[scale=1,valign=c]{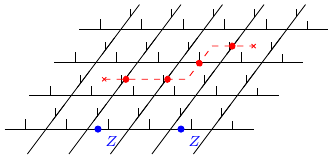} 
\label{eq:creata_anyon}
\end{equation}
Consequently, anyonic correlators can be evaluated through $\sigma_A$ and $\sigma_{\bar A}$, and also therefore encoded in $\rho_{\partial A}$.
For instance,
\begin{equation}
    \begin{aligned}
        \lrangle{e_ie_j}&=\Tr[\sigma_A^T Z_i Z_j\sigma_{\bar A}]=\Tr[\rho_{\partial A}Z_iZ_j]=0,\\
         \lrangle{m_im_j}&=\Tr\left[\sigma_A^T \left(\prod_{k=i}^jX_k\right)\sigma_{\bar A}\right]=\Tr\left[\rho_{\partial A}\prod_{k=i}^j X_k\right]=0,
    \end{aligned}
    \label{eq:ee_corre}
\end{equation}
The vanishing correlation $\lrangle{a_ia_j}=0$ confirms that these string operators creates a state orthogonal to the ground state, indicating that neither anyon species condenses. 
For a 2D gapped system, the simultaneous non-condensation of both $e$ and $m$ anyons is the hallmark of the deconfined phase of a $\ZZ_2$ gauge theory.

This fixed-point result realizes our definition of SW-SSB for 1D mixed states in the main text.
As established in Eq.~\eqref{eq:fp_strongz2}, $\rho_{\partial A}$ possesses a strong $\ZZ_2$ symmetry generated by $\prod_j X_j$. 
The two defining correlators of SW-SSB — the charged-operator correlator $\Tr(\rho_{\partial A} Z_i Z_j)$ and the domain-wall correlator $\Tr(\rho_{\partial A} \prod_{k=i}^j X_k)$ — both vanish at the fixed point (Eq.~\eqref{eq:ee_corre}), signaling the spontaneous breaking of the strong $\ZZ_2$ symmetry to a weak one.

\subsubsection{General relation with SW-SSB}
The fixed-point derivation above relied on the exact solvability of Kitaev's toric code. We now show that the connection between deconfinement and SW-SSB is in fact a generic feature of the entire phase, originating from the emergent one-form symmetry that defines topological order.

The key insight is that a generic state in toric code topological order can be realized by PEPS with $\ZZ_2$ symmetry solely acting on virtual legs~\cite{schuch2010peps}:
\begin{equation}
    V = \adjincludegraphics[scale=1,valign=c]{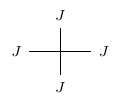}
,~ E =\adjincludegraphics[scale=1,valign=c]{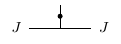}
\label{eq:igg_beyond_fp}
\end{equation}
At the fixed point, $J = X$ and Eq.~\eqref{eq:igg_beyond_fp} reduces to a subset of the fixed-point constraints in Eq.~\eqref{eq:toric_code_local_tensor_constraint}.  
Here, we point out that this $\ZZ_2$ symmetry is not a fine-tuned property of the fixed point; it is a robust feature of the entire toric code phase~\cite{schuch2010peps}.
Physically, it originates from the $m$-particle one-form symmetry.

Equipped with this virtual $\ZZ_2$ symmetry, anyon excitations are easily constructed: a pair of $m$-particles is generated by an open $J$-string, and the $e$-particle can be identified as local operator on virtual legs anticommute with $J$.
These operators are diagrammatically similar as those illustrated in Eq.~\eqref{eq:creata_anyon}.

The $\ZZ_2$ virtual-leg symmetry further translates into a $\ZZ_2$ strong symmetry for the state $\rho_{\partial A}$: $\rho_{\partial A}=(\prod_j J_j)\rho_{\partial A}$, as Eq.~\eqref{eq:igg_beyond_fp} guarantees that both $\sigma_A$ and $\sigma_{\bar A}$ are invariant under $\prod_j J_j$.
By further adding the non-condensation condition, we conclude~\cite{haegeman2015shadows,Duivenvoorden2017Entanglement}:
\begin{equation}
    \begin{aligned}
        \lrangle{e_ie_j}=\Tr[\sigma_A^T \wt{Z}_i \wt{Z}_j\sigma_{\bar A}]\sim\Tr[\rho_{\partial A}\wt{Z}_i\wt{Z}_j]\sim\ee^{-\abs{i-j}/\xi_e}\to0\\
         \lrangle{m_im_j}=\Tr[\sigma_A^T (\prod_{k=i}^jJ_k)\sigma_{\bar A}]\sim\Tr[\rho_{\partial A}\prod_{k=i}^j J_k]\sim\ee^{-\abs{i-j}/\xi_m}\to0\\
    \end{aligned}
    \label{eq:ee_mm_corre}
\end{equation}
Hence, $\rho_{\partial A}$ represents a 1D mixed state exhibiting SW-SSB.

\paragraph{Higgs phase.}
We now extend the analysis to the Higgs phase, where $e$-particles condense.
The behavior of $m$-particle correlators requires closer inspection.
In this phase, $m$-particles are confined: the energy required to separate a pair grows linearly with distance (dynamical confinement).
This forces the spatial wavefunction of the pair to collapse~\cite{Huxford2023deformedtoriccode}.
The relevant correlators in the Higgs phase are then:
\begin{equation}
    \begin{aligned}
         & \lrangle{m_im_j|m_im_j}=\Tr[\sigma_A^T (\prod_{k=i}^jJ_k)\sigma_{\bar A}(\prod_{k=i}^jJ_k)]=\Tr[\sqrt{\rho_{\partial A}}\prod_{k=i}^j J_k\sqrt{\rho_{\partial A}}\prod_{k=i}^j J_k]\sim\ee^{-\abs{i-j}/\xi_m^{(2)}}\\
         &\lrangle{e_ie_j}=\Tr[\sigma_A^T \wt{Z}_i \wt{Z}_j\sigma_{\bar A}]=\Tr[\rho_{\partial A}\wt{Z}_i\wt{Z}_j]\sim O(1)\\
    \end{aligned}
    \label{eq:ee_corre_higgs}
\end{equation}
which indicates that $\rho_{\partial A}$ realizes a conventional SSB phase.

\paragraph{Confined phase.}
Exchanging the roles of $e$ and $m$ in the Higgs phase yields the confined phase, where $m$-particles condense while $e$-particles are confined. 
The corresponding correlators are:
\begin{equation}
    \begin{aligned}
         & \lrangle{e_ie_j|e_ie_j}=\Tr[\sigma_A^T \wt{Z}_i\wt{Z}_j\sigma_{\bar A}\wt{Z}_i^\dagger\wt{Z}_j^\dagger]=\Tr[\sqrt{\rho_{\partial A}}\wt{Z}_i\wt{Z}_j\sqrt{\rho_{\partial A}}\wt{Z}_i^\dagger\wt{Z}_j^\dagger]\sim\ee^{-\abs{i-j}/\xi_e^{(2)}},\\
         &\lrangle{m_im_j}=\Tr[\sigma_A^T(\prod_{k=i}^j J_k)\sigma_{\bar A}]=\Tr[\rho_{\partial A}(\prod_{k=i}^j J_k)]\sim O(1).
    \end{aligned}
    \label{eq:mm_corre_confine}
\end{equation}
Therefore, $\rho_{\partial A}$ realizes a strong $\ZZ_2$ symmetric phase.

In conclusion, Eqs.~\eqref{eq:ee_corre_higgs}--\eqref{eq:mm_corre_confine} complete the dictionary between the three mixed-state phases (SW-SSB, SSB, symmetric) and the three known phases of $\ZZ_2$ gauge theory (deconfined, $e$-condensed, $m$-condensed).

\subsection{Discussion on corner entanglement cut}
In the main text and preceding sections, our calculations assumed a straight boundary for subregion $A$.
This approach extends naturally to entanglement cuts with corners, which is inevitable to cut a open disk subregion.

Following the main text, we decompose the total Hamiltonian into three parts:
\begin{equation}
    H=-\sum_{p}\prod_{i\in p}Z_i-\sum_{v}\prod_{i\in v}X_i
        \equiv H_A+H_{\bar A}+H_{\partial A\cup\partial \bar A}
\end{equation}
where $H_{A/\bar{A}}$ contains the interaction terms supported strictly within $A$ and $\bar{A}$, respectively. 
We aim to determine the boundary algebra of $A$---namely, the set of operators commuting with $H_A$---as these specify the zero-mode space of $H_A$.
As shown in the main text, when $\partial A$ is locally a straight line, the boundary algebra is generated by
\begin{equation}
    \{X^A_{2k},~Z^A_{2k}Z^A_{2k+1}Z^A_{2k+2}\}_{k}
\end{equation}
This algebraic structure dictates that each $X$-term anticommutes with its two adjacent three-$Z$ terms.

Near a corner, the generators of the boundary algebra take the form:
\begin{equation}
    \adjincludegraphics[scale=1,valign=c]{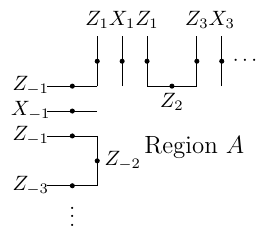}
\end{equation}
where operators sharing the same subscript reside on the same bond of the square lattice.
Evidently, this algebra preserves the same structural relation: a single $X$-operator anticommutes with its two nearest $Z$-terms.
The only distinction is that while the $Z$-terms on the smooth boundary consist of three $Z$ operators, those located exactly at the corner consist of only two. 

Consequently, this boundary algebra admits the exact same representation mapped to a 1D spin chain.
In the ground state, the interior of $A$ is free of anyon excitations, trivially guaranteeing that the two Wilson loops have eigenvalues $W_e=W_m=1$.

The corresponding PEPS analysis for cornered cuts directly parallels the Hamiltonian derivation above; hence, we omit the details here.

\section{Comparing different definitions of SW-SSB phases}\label{app:diff_def_sw_ssb}
In this section, we compare two complementary definitions of the $\ZZ_2$ SW-SSB phase: the Wightman/fidelity correlator of charged operators, and our proposed definition based on the disorder parameter (domain-wall correlator). We highlight the hidden assumptions underlying their equivalence and provide counterexamples where they disagree.

\paragraph{SW-SSB from the canonical purification.}
It has been conjectured that mixed-state phases of $\rho$ are closely related to pure-state phases of its canonical purification $\kket{\sqrt{\rho}}$.
For a mixed state $\rho$ with $\ZZ_2$ strong symmetry generated by $\prod_j X_j$, the purification $\kket{\sqrt{\rho}}$ possesses a $\ZZ_2\times \ZZ_2$ symmetry, acting on the ket and bra degrees of freedom, respectively.
In this framework, the SW-SSB phase of $\rho$ corresponds to spontaneous breaking of $\ZZ_2\times\ZZ_2$ down to the diagonal $\ZZ_2$ subgroup in the purified state.
This diagonal SSB phase of $\kket{\sqrt{\rho}}$ is characterized by long-range correlations of operators charged under $\ZZ_2\times\ZZ_2$ but neutral under the diagonal subgroup, while diagonal-charged operators remain short-range correlated.
Concretely,
\begin{equation}
\begin{aligned}
\Tr(\sqrt{\rho} Z_iZ_j \sqrt{\rho}Z_iZ_j)&=\bbra{\sqrt{\rho}} (Z_i\otimes Z_i)\cdot (Z_j\otimes Z_j)\kket{\sqrt{\rho}}\sim O(1)~,\\
\Tr(\rho Z_iZ_j)&=\bbra{\sqrt{\rho}} (Z_i\otimes\hat{1})\cdot (Z_j\otimes\hat{1})\kket{\sqrt{\rho}}\sim\exp(-\abs{i-j}/\xi)
\end{aligned}
\label{eq:sw_ssb_wightman_corr}
\end{equation}
The first line is the Wightman correlator~\cite{Liu2025diagnosingSWSSB,weinstein2024swssb}, proposed as a diagnostic for spontaneous breaking of the strong $\ZZ_2$ symmetry. It has been shown to be equivalent to the original fidelity-based definition of SW-SSB~\cite{Lessa2025SWSSB}.

\paragraph{Alternative definition via the disorder parameter.}
In practice, the Wightman correlator is challenging to measure — both experimentally and numerically — owing to its nonlinear dependence on $\rho$.
This motivates an alternative definition based on quantities linear in $\rho$.
Inspired by the characterization of standard SSB phases via the disorder parameter~\cite{Kadanoff1971disorder}, we propose that strong symmetry breaking can be diagnosed by volume-law scaling of the disorder parameter associated with the strong symmetry:
\begin{equation}
\Tr(\rho X_A)=\bbra{\sqrt{\rho}}X_A\otimes\hat{1}\kket{\sqrt{\rho}}\sim\exp(-\beta V_A)
\label{eq:sw_ssb_disorder_param}
\end{equation}
Here $X_A=\prod_{j\in A} X_j$ for a large subregion $A$ creates domain walls at $\partial A$, $V_A$ is the volume of $A$, and $\beta$ is a non-universal constant.
Table~\ref{tab:mixed_phases_def} summarizes the three phases of mixed states with $\ZZ_2$ strong symmetry under both diagnostics.

\begin{table}[htpb]
\centering
\renewcommand{\arraystretch}{1.8}
\begin{tabular}{|p{0.2\textwidth}|p{0.2\textwidth}|p{0.2\textwidth}|p{0.2\textwidth}|}
\hline
\centering Mixed state phases & \centering $\Tr(\rho Z_iZ_j)$ & \centering $\Tr(\sqrt{\rho}Z_iZ_j\sqrt{\rho}Z_iZ_j)$ & \centering $\Tr(\rho X_A)$ \tabularnewline
\hline
\centering Strong Symmetric & \centering $\exp(-\abs{i-j}/\xi)$ & \centering $\exp(-\abs{i-j}/\xi')$ & \centering $\exp(-\alpha S_{\partial A})$ \tabularnewline
\hline
\centering SW-SSB & \centering $\exp(-\abs{i-j}/\xi)$ & \centering $O(1)$ & \centering $\exp(-\beta V_A)$ \tabularnewline
\hline
\centering SSB & \centering $O(1)$ & \centering $O(1)$ & \centering $\exp(-\beta V_A)$ \tabularnewline
\hline
\end{tabular}
\caption{Behavior of charge correlator, Wightman correlator, and disorder parameter across the three mixed-state phases with $\ZZ_2$ strong symmetry.}
\label{tab:mixed_phases_def}
\end{table}

\paragraph{When do the two definitions agree?}
The equivalence between the Wightman correlator (Eq.~\eqref{eq:sw_ssb_wightman_corr}, first line) and the volume-law disorder parameter (Eq.~\eqref{eq:sw_ssb_disorder_param}) rests on two implicit assumptions:
\begin{enumerate}
    \item $\kket{\sqrt{\rho}}$ shares the qualitative properties of a ground state of a local Hamiltonian, so that it is meaningful to speak of its quantum phase.\label{enu:local_ham}
    \item In such a local-Hamiltonian ground state, long-range order of a charged operator is a sufficient and necessary condition for volume-law scaling of the corresponding disorder parameter.\label{enu:vol_law_disorder_param}
\end{enumerate}
Both assumptions can fail, as we now illustrate. Below we present two examples where the Wightman-based and disorder-parameter-based definitions of SW-SSB are inequivalent.

\paragraph{Example 1: violation of locality (Assumption~\ref{enu:local_ham}).}
Consider the mixed state from Ref.~\cite{Lessa2025SWSSB},
\begin{equation}
    \rho=\frac{1}{2}\ket{++\cdots}\bra{++\cdots}+\frac{1}{2^{L+1}}(\hat{1}+\prod_{j=1}^L X_j).
\end{equation}
Here the charge correlator vanishes, $\lrangle{Z_i Z_j}=0$, while the Wightman correlator is long-range ordered: $\bbra{\sqrt{\rho}}(Z_i\otimes{Z}_i)\cdot ( Z_j\otimes{Z}_j)\kket{\sqrt{\rho}}\sim O(1)$.
The Wightman correlator thus diagnoses SW-SSB.
However, the disorder parameter saturates at a constant, $\lrangle{X_A}=\frac{1}{2}$, violating volume-law scaling.
The discrepancy arises because $\kket{\rho}$ is not a ground state of any local Hamiltonian, so Assumption~\ref{enu:local_ham} does not hold.

\paragraph{Example 2: violation of the order--disorder connection (Assumption~\ref{enu:vol_law_disorder_param}).}
The decohered Ising model provides a counterexample to Assumption~\ref{enu:vol_law_disorder_param}~\cite{sahay2025enforcedgaplessnessstatesexponentially}.
Starting from the fully polarized state $\ket{+++\dots}$ and applying a $\ZZ_2$ strong symmetric quantum channel $\mathcal{E}=\prod_{\lrangle{ij}}\EE_{ij}$ with
\begin{equation}
    \EE_{ij}(\rho)=(1-p)\rho+p Z_iZ_j\cdot\rho\cdot Z_i Z_j,
\end{equation}
the Wightman correlator exhibits a transition\cite{Lessa2025SWSSB}:
\begin{equation}
    \Tr(\sqrt{\rho}Z_iZ_j\sqrt{\rho}Z_iZ_j)\sim
    \begin{cases}
        \exp(-\abs{i-j}/\xi), & p<p_c\\
        O(1) , & p>p_c.
    \end{cases}
\end{equation}
To compute the disorder parameter, we rewrite $\rho$ in the bond-variable representation,
\begin{equation}
    \rho = [p(1-p)]^{N_b/2}\cdot \sum_{\{\mu_b\}}\ee^{\beta\sum_b\mu_b}\ket{\{x_v=\prod_{b\in v}\mu_{b}\}}\bra{\{x_v\}},
\end{equation}
where $\mu_{ij}=-1$ indicates that bond $b=\lrangle{ij}$ has been acted on by $Z_i Z_j$, and $\ee^{2\beta}=\frac{1-p}{p}$ (equivalently $\tanh\beta=1-2p$). The disorder operator is diagonal in this basis, giving
\begin{equation}
\begin{aligned}
    \lrangle{X_A}&=\frac{1}{Z_0}\sum_{\{\mu_b\}}\left(\prod_{b'\in \partial A}\mu_{b'}\right)\ee^{\beta\sum_{b}\mu_b}\\
    &=\frac{1}{Z_0}\left( \prod_{b\in\partial A}\sum_{\mu_b=\pm1}\mu_b\,\ee^{\beta\mu_b} \right)\cdot \left(\prod_{b\notin \partial A} \sum_{\mu_b=\pm1}\ee^{\beta\mu_b} \right)\\
    &=(\tanh\beta)^{S_{\partial A}}
    =(1-2p)^{S_{\partial A}},
\end{aligned}
\end{equation}
with $Z_0=\sum_{\{\mu_b\}}\ee^{\beta\sum_b\mu_b}=(2\cosh{\beta})^{N_b}$. The disorder parameter obeys area-law scaling for all $p$, even when the Wightman correlator is long-range ordered. Thus, the two definitions disagree: the Wightman correlator signals SW-SSB for $p>p_c$, while the disorder parameter never does.

These examples demonstrate that the Wightman/fidelity definition and the disorder-parameter definition of SW-SSB are not universally equivalent.

\section{Kramers-Wannier transformation in the mixed-state phases and the bulk $e$-$m$ duality}\label{app:dual_peps_kw}
The toric code possesses a electromagnetic ($e$-$m$) duality: exchanging $e$ and $m$-anyons maps the deconfined phase to itself.
In the main text, we identified the 1D mixed state $\rho_{\partial A}$ as the central object encoding the bulk topological order, and we established that deconfinement corresponds to SW-SSB of $\rho_{\partial A}$.
A natural question arises: how does the bulk $e$-$m$ duality manifest in the 1D mixed-state description?
In this section, we show that it descends to a Kramers-Wannier (KW) self-duality of $\rho_{\partial A}$.
We provide the explicit mapping between the two representations of the boundary algebra, discuss its physical interpretation, and show how it acts on the phase diagram.

\paragraph{Boundary algebra and the 1D representation.}
We begin by recalling the Hamiltonian decomposition from the main text. The toric code Hamiltonian on a bipartitioned system is
\begin{equation}
    H=-\sum_{p}\prod_{i\in p}Z_i-\sum_{v}\prod_{i\in v}X_i
        \equiv H_A+H_{\bar A}+H_{\partial A\cup\partial \bar A}
\end{equation}
where $H_{A/\bar{A}}$ contains interaction terms supported strictly within $A/\bar{A}$. The zero-mode space of $H_A$ is governed by the following boundary algebra:
\begin{equation}
    \{X^A_{2k},~Z^A_{2k}Z^A_{2k+1}Z^A_{2k+2}\}_{k=1,2,\dots,L}
\end{equation}

\paragraph{The $\tau$ representation.}
A 1D spin chain provides a faithful representation of the boundary algebra:
\begin{equation}
     \tau^{x}_{k}\equiv  X^A_{2k}~,\quad
    \tau^{z}_k\tau^{z}_{k+1}\equiv  Z^A_{2k}Z^A_{2k+1}Z^A_{2k+2}~
\end{equation}
In this representation, the anyon loops map to
\begin{equation}
    W_m = \prod_{k=1}^L X_{2k}^A = \prod_{k=1}^L \tau_k^x,\quad W_e = \prod_{k=1}^L Z_{2k+1}^A = \tau_1^z \tau_{L+1}^z.
\end{equation}
Physically, the $m$-loop becomes a global $\ZZ_2$ symmetry of the 1D chain (generated by $\prod_k \tau_k^x$), while the $e$-loop reduces to a boundary condition ($\tau_1^z \tau_{L+1}^z$).

\paragraph{The $\mu$ representation and KW duality.}
The choice of 1D representation is not unique. An equally valid alternative is
\begin{equation}
     \mu^{z}_k\mu^{z}_{k+1} \equiv  X^A_{2k}~,\quad
   \mu^{x}_{k+1}\equiv  Z^A_{2k}Z^A_{2k+1}Z^A_{2k+2},
\end{equation}
under which the anyon loops exchange roles:
\begin{equation}
    W_m = \prod_{k=1}^L X_{2k}^A= \mu_1^z \mu_{L+1}^z,\quad W_e = \prod_{k=1}^L Z_{2k+1}^A  = \prod_{k=1}^L \mu_k^x.
\end{equation}
In the $\mu$ representation, the $e$-loop becomes the global symmetry and the $m$-loop becomes the boundary condition — precisely the opposite of the $\tau$ representation.

The mapping between these two representations is
\begin{equation}
    \tau^{x}_{k}\to \mu^{z}_k\mu^{z}_{k+1},\quad \tau^{z}_k\tau^{z}_{k+1}\to\mu^{x}_{k+1},
\end{equation}
which is the celebrated Kramers-Wannier (KW) transformation for a 1D spin chain with a $\ZZ_2$ symmetry~\cite{Kadanoff1971disorder}.

\paragraph{Self-duality of $\rho_{\partial A}$ under KW.}
We now apply the KW transformation to $H_{\partial A \cup \partial \bar A}$. 
In the $\tau$ representation,
\begin{equation}
    H_{\partial A\cup\partial \bar A}=-\sum_{k=1}^{L}(\tau^{x}_{k}\bar{\tau}^{x}_{k}+\tau^{z}_{k}\tau^{z}_{k+1}\bar{\tau}^{z}_{k}\bar{\tau}^{z}_{k+1}).
\end{equation}
The $\mu$ represenetation gives
\begin{equation}
    H_{\partial A\cup\partial \bar A} \;\longrightarrow\; -\sum_{k=1}^{L}(\mu^{z}_{k}\mu^{z}_{k+1}\bar{\mu}^{z}_{k}\bar{\mu}^{z}_{k+1}+\mu^{x}_{k}\bar{\mu}^{x}_{k}),
\end{equation}
which relates to $\tau$ representation by KW transformation, and the form of the coupling is preserved. 
This demonstrates that the pure-state diagonal SSB pattern, $\ZZ_2^a \times \ZZ_2^b \to \ZZ_2^{ab}$, is self-dual under KW.
Since this pure-state phase is precisely the purification of $\rho_{\partial A}$, the SW-SSB mixed-state phase inherits the same self-duality, where KW transformation exchange the strong-symmetry generator with the boundary condition.

The SW-SSB phase — defined by the simultaneous exponential decay of \emph{both} correlators — is manifestly self-dual.
By contrast, KW maps the symmetric phase (where the domain wall is long-range ordered) to the SSB phase (where the charged operator is long-range ordered), and vice versa.
This is summarized in Table~\ref{tab:pure_state_diag}.

\begin{table}[htpb]
\centering
\renewcommand{\arraystretch}{1.8}
\begin{tabular}{|p{0.2\textwidth}|p{0.2\textwidth}|p{0.2\textwidth}|p{0.2\textwidth}|p{0.2\textwidth}|}
\hline
\centering Pure-state phases & \centering $G\to H$ pattern & \centering Correlators & \centering Mixed-state phases & \centering 2D bulk phases \tabularnewline
\hline
\centering Diagonal SSB & \centering $\ZZ_2^a\times\ZZ_2^b\to\ZZ_2^{ab}$ & \centering $\lrangle{Z^a Z^b}\neq0,\lrangle{Z^{a/b}}=0$ & \centering SW-SSB & \centering Toric code phase \tabularnewline
\hline
\centering Symmetric & \centering $\ZZ_2^a\times\ZZ_2^b\to\ZZ_2^a\times\ZZ_2^b$ & \centering $\lrangle{Z^a},\lrangle{Z^b},\lrangle{Z^{a}Z^b}=0$ & \centering Symmetric & \centering Confined phase \tabularnewline
\hline
\centering SSB & \centering $\ZZ_2^a\times\ZZ_2^b\to0$ & \centering $\lrangle{Z^a},\lrangle{Z^b},\lrangle{Z^{a}Z^b}\neq0$ & \centering SSB (strong-to-none) & \centering Higgs phase \tabularnewline
\hline
\end{tabular}
\caption{Correspondence among 1D pure-state phases (at $\partial A\cup\partial\bar{A}$), 1D mixed-state phases of $\rho_{\partial A}$, and 2D topological phases.
    Under the KW duality, the ``Symmetric'' and ``SSB'' rows are exchanged, while the ``Diagonal SSB'' (SW-SSB) row is self-dual.
    The correlator notation $\lrangle{Z^{a/b}}$ refers to the charged operator in the $\ZZ_2^a$ or $\ZZ_2^b$ sector of the purified state.
}
\label{tab:pure_state_diag}
\end{table}

\section{Details of Numerical Studies}\label{app:peps_rvb_numerics}
This section contains two numerical studies.
The first verifies the full three-phase dictionary of the pure--mixed correspondence (deconfined $\leftrightarrow$ SW-SSB, confined $\leftrightarrow$ strong-symmetric, Higgs $\leftrightarrow$ SSB) using a deformed toric code model.
We compute both the linear correlators (charge and domain-wall) and the Wightman correlators from the MPDO representation of $\rho_{\partial A}$, confirming that they exhibit the expected behavior across the phase diagram.
The second study computes the spin-rotation disorder parameter for the NN-RVB state on the kagome lattice using tensor renormalization group (TRG) contraction, verifying the predicted cusp at $\theta = \pi$ that signals spin-$\frac{1}{2}$ spinons.

\subsection{Numerical verification of SW-SSB in the RDM of deformed toric code model}
The correspondence between 2D topological phases and 1D mixed-state phases can be illustrated numerically in the deformed toric code model
\begin{align}
    |\psi\rangle\propto\prod_i\exp{\left(\frac{\beta_x}{4}X_i+\frac{\beta_z}{4}Z_i\right)}|TC\rangle,
    \label{eq:deformed_TC}
\end{align}
where $\ket{TC}$ is the ground state of the toric code model, whose PEPS form is given in supplemental section \ref{app:Fixed_pt_PEPS}.
In the PEPS representation, this deformation modifies the local $E$ tensor of Eq.~\eqref{eq:TC_PEPS_tensor} at each link.
As discussed in supplemental section \ref{app:RDM-entanglement-space}, the RDM at the entanglement cut is $\rho_{\partial A}=\sqrt{\sigma_A^T}\,\sigma_{\bar A}\,\sqrt{\sigma_A^T}$.

The choice of entanglement cut is not unique. For the deformed toric code, we can choose the cut along the diagonal direction that passes through the $V$ tensors (using the same notation as Sec.~\ref{app:Fixed_pt_PEPS}). This cut gives $\sigma_A^T = \sigma_{\bar A}=\sqrt{\rho_{\partial A}}$, which simplifies the computation of $\rho_{\partial A}$.
Specifically, the $V$ tensor solving from Eq.~\eqref{eq:toric_code_local_tensor_constraint} can be splited into two 3-leg tensors in two ways, 
\begin{equation}
    V =\adjincludegraphics[scale=1,valign=c]{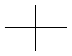}
= \adjincludegraphics[scale=1,valign=c]{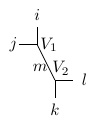}=
\adjincludegraphics[scale=1,valign=c]{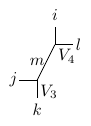},
\end{equation}
satisfying
\begin{equation}
     V_{1,ijm}=V_{1,jim}, V_{2,klm}=V_{2,lkm}, V_{3,jkm}=V_{3,kjm}, V_{4,ilm}=V_{4,lim}, V_{1,ijm}=V_{2,lkm}, V_{3,jkm}=V_{4,ilm},
\end{equation}
and 
\begin{equation}
    V_1 =\adjincludegraphics[scale=1,valign=c]{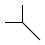}
    = \adjincludegraphics[scale=1,valign=c]{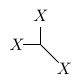},\ \ 
    V_2 =\adjincludegraphics[scale=1,valign=c]{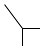}
    = \adjincludegraphics[scale=1,valign=c]{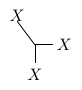},\ \ 
    V_3 =\adjincludegraphics[scale=1,valign=c]{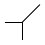}
    = \adjincludegraphics[scale=1,valign=c]{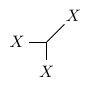},\ \ 
    V_4 =\adjincludegraphics[scale=1,valign=c]{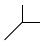}
    = \adjincludegraphics[scale=1,valign=c]{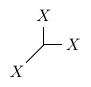}.
\end{equation}
And the $E$ tensor solving from Eq.~\eqref{eq:toric_code_local_tensor_constraint} is symmetric under the permutation of the two virtual legs. The local tensor of the deformed toric code model can be redefined on each even plaquette,
\begin{equation}
    T=\adjincludegraphics[scale=1,valign=c]{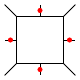},
\end{equation}
where the red dots denote the action of the operator $\exp\left(\frac{\beta_x}{4}X+\frac{\beta_z}{4}Z\right)$ on the physical legs. This tensor is symmetric under the reflection along the two diagonal directions. Therefore, entanglement cut along the diagonal direction results in $\sigma_A^T = \sigma_{\bar A}=\sqrt{\rho_{\partial A}}$.

In the thermodynamic limit, $\sigma_A$($\sigma_{\bar A}$) is the left(right) dominant eigenvector of the transfer matrix, which can be approximated by the variational uniform MPS (VUMPS) \cite{Zauner_Stauber_2018VUMPS} algorithm. 
The anyon correlators are calculated with the boundary MPS, which can further be interpreted as the charge and domain-wall correlators of the 1D mixed state $\rho_{\partial A}$:
\begin{equation}
    \begin{aligned}
        \lrangle{e_ie_j}&=\Tr[\sigma_A^T Z_i Z_j\sigma_{\bar A}]=\Tr[\rho_{\partial A}Z_iZ_j],\\
         \lrangle{m_im_j}&=\Tr\left[\sigma_A^T \left(\prod_{k=i}^jX_k\right)\sigma_{\bar A}\right]=\Tr\left[\rho_{\partial A}\prod_{k=i}^j X_k\right],
    \end{aligned}
    \label{eq:ee_corre_numeric}
\end{equation}
Further, 
\begin{equation}
\begin{aligned}
    &\lrangle{e_ie_j|e_ie_j}=\Tr[\sigma_A^T Z_iZ_j\sigma_{\bar A}Z_i^\dagger Z_j^\dagger]=\Tr[\sqrt{\rho_{\partial A}}Z_iZ_j\sqrt{\rho_{\partial A}}Z_i^\dagger Z_j^\dagger], \\
    &\lrangle{m_im_j|m_im_j}=\Tr[\sigma_A^T (\prod_{k=i}^jX_k)\sigma_{\bar A}(\prod_{k=i}^jX_k)]=\Tr[\sqrt{\rho_{\partial A}}\prod_{k=i}^j X_k\sqrt{\rho_{\partial A}}\prod_{k=i}^j X_k],   
\end{aligned}
\end{equation}
identified as the Wightman correlators in the purified state $\kket{\sqrt{\rho_{\partial A}}}$\cite{Liu2025diagnosingSWSSB,weinstein2024swssb}.

\begin{figure}
    \centering
    \includegraphics[width=0.55\linewidth]{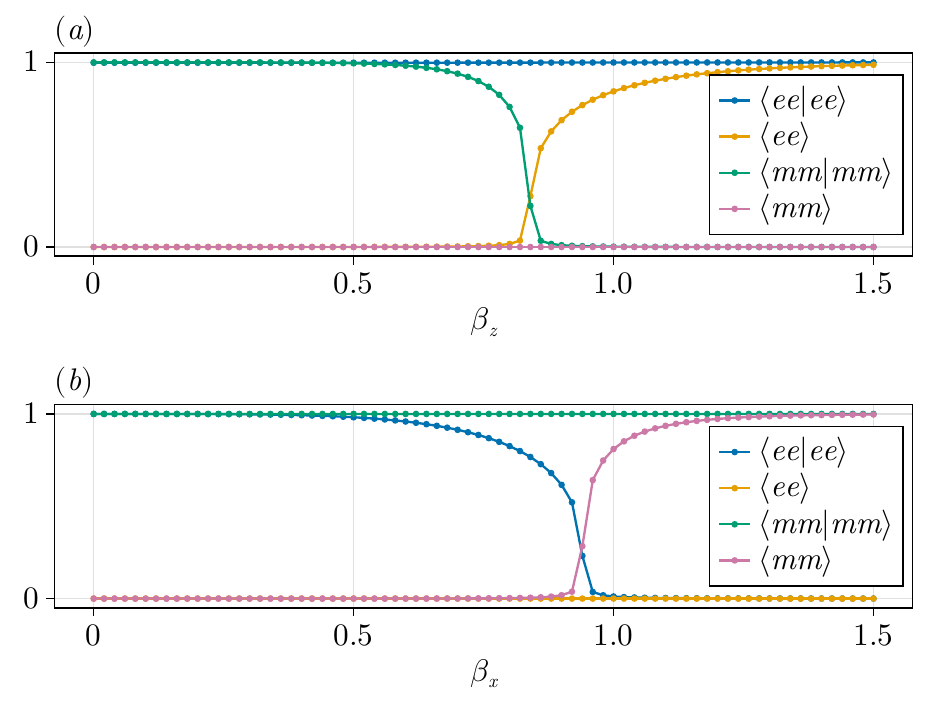}
    \caption{(a)/(b): Correlation functions of deformed toric code wave function with parameter $\beta_z$/$\beta_x$.
        Here, $\beta_x/\beta_z$ are set to 0.1 for (a)/(b).
        Deconfined phase lies in the region $\beta_{z,x}<\beta_{z,x}^c$, while $e$/$m$ condenses if $\beta_{z,x}>\beta_{z,x}^c$ .
        We set cut-off bond dimension $\chi=16$ when optimizing boundary MPS.
    All correlation functions are presented in the thermodynamic limit.}
    \label{fig:num_tc}
\end{figure}

Here, we plot these correlation for different $\beta_x$ and $\beta_z$ in Fig. \ref{fig:num_tc}. 
Three regimes are clearly visible: (i) For $\beta_{z,x} < \beta_{z,x}^c$, both $e$ and $m$ correlators decay exponentially — this is the deconfined phase, corresponding to SW-SSB of $\rho_{\partial A}$.
(ii) For $\beta_x=0,~\beta_z>\beta_z^c$, the $e$-anyon condenses, while $m$-anyon correlator remains short-range — the Higgs phase, corresponding to SSB.
(iii) For $\beta_z=0,~\beta_x > \beta_x^c$, the $m$-anyon condenses — the confined phase, corresponding to strong-symmetric.


\subsection{Disorder parameter of RVB state on the kagome lattice}
\begin{figure} 
    \centering
    \adjincludegraphics[scale=1,valign=c]{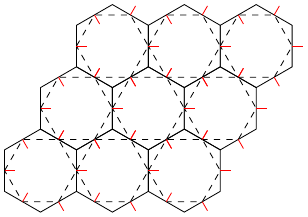}
    \caption{The PEPS form of NN-RVB state. Black dashed lines are links of a kagome lattice. Black solid lines are legs in the virtual spaces. Red dashed lines are legs in the physical spaces.}
    \label{fig:kagome_peps}
\end{figure}

The behavior of the disorder parameter in the gapped $\ZZ_2$ spin liquid phase is tested in the spin-$\frac12$ NN-RVB state on the kagome lattice~\cite{pepsrvb2012}.
The PEPS form is shown in Fig. \ref{fig:kagome_peps}, with details list in the following.
The virtual leg is a qutrit space, with basis $\{\ket{0}, \ket{1}, \ket{2}\}$.
For each triangle, we assign a 3-leg tensor, where
\begin{equation}
|\epsilon_{\triangle}\rangle=\sum_{i,j,k=0}^2\epsilon_{ijk}\ket{ijk}+\ket{222},
\end{equation}
where $\epsilon_{ijk}$ is the antisymmetric tensor with $\epsilon_{012}=1$.
The site tensor is defined as 
\begin{equation}
    P=\ket{0}(\bra{02}+\bra{20})+\ket{1}(\bra{12}+\bra{21}),
\end{equation}
The NN-RVB state is then obtained by contracting all virtual legs:
\begin{equation}
    \ket{\psi} = \prod_i P_i\bigotimes_{\triangle}\ket{\epsilon_{\triangle}},
\end{equation}

The $U(1)$ spin rotation symmetry along $z$ axis is given by 
\begin{equation}
    U(\theta) = \exp\left( i\frac{\theta}{2} Z\right),
\end{equation}
inducing gauge transformation on the virtual space
\begin{equation}
W(\theta) =\left(
    \begin{array}{cc}
       \exp\left( i\frac{\theta}{2} Z\right)  &  \\
         & 1
    \end{array}
    \right).
\end{equation}
Note that the PEPS is symmetric under a $\ZZ_2$ virtual symmetry 
\begin{equation}
J =\left(
    \begin{array}{ccc}
     -1 & & \\
       & -1 &  \\
       & & 1
    \end{array}
    \right).
\end{equation}
which is identified as $2\pi$ spin rotation in the virtual leg: $W(2\pi) = J$.
Physically, $J$ is the $m$-string, and $W(2\pi) = J$ is the symmetry fractionalization condition where $e$-anyon carries spin-$\frac12$.

The disorder parameter $\langle U_A(\theta)\rangle=\bra{\psi}U_A(\theta)\ket{\psi}$ can be calculated using the TRG \cite{Gu2008TRG} method. The TRG contraction was performed with bond dimension $\chi_{\rm TRG} = 32$, and we take the $16\times16$ square lattice with PBC. The function $f(\theta) = -\ln|\langle U_A(\theta)\rangle| / L$ was extracted from the logarithm of the contracted network normalized by the boundary length $L$. For finite $L$, $f(\theta)$ become more and more singular as $L$ increases, but is always smooth at $\theta =\pi$, and a cusp will appear as $L\to\infty$, as shown in Fig.~1 of the main text. This is a robust feature that persists upon increasing $\chi_{\rm TRG}$, confirming that it is not a finite-bond-dimension artifact. The discontinuity in the derivative $f'(\theta)$ at $\theta = \pi$ is the numerical signature of the eigenvalue crossing derived analytically in Sec.~\ref{app:z2_set_imo}.

\section{Correlators and virtual leg symmetries in MPDO}\label{app:mps_mpdo_correlator}
The RDM at the entanglement cut of a 2D topologically ordered phase realizes a 1D SW-SSB phase, which is diagnosed by the long-range behavior of the charged-operator correlator and the domain-wall correlator. 
In this section, we show how these correlators are controlled by virtual-leg symmetries of the underlying tensor network representation.
We first review the pure-state MPS formalism for the 1D $\ZZ_2$ Ising chain, where the key concepts — virtual leg symmetries and transfer matrix degeneracy — are most transparent.
We then generalize to the MPDO formalism and show that the $\ZZ_2$ SW-SSB phase is characterized by an analogous virtual-leg symmetry. 

\subsection{$\ZZ_2$ Ising chain and MPS}
We consider the transverse field Ising model in 1+1D, where
\begin{equation}
    H=\sum_{j=1}^L -J Z_j Z_{j+1} -h X_j
    \label{eq:tfim_ham}
\end{equation}
with periodic boundary condition imposed.
The system exhibits a global $\ZZ_2=\{1,g\}$ symmetry, generated by $\prod_j X_j$.
For $J>h$, the system is in the $\ZZ_2$ SSB phase, characterized by a two-fold degeneracy of ground states in the thermodynamic limit, which are two cat states with distinct $\ZZ_2$ quantum number $\pm1$.
In these two cat states, while $\ZZ_2$-charged operators exhibit vanishing expectation values, they are long-range correlated in the thermodynamic limit $L\to\infty$:
\begin{equation}
\lrangle{Z_1}=0~,\quad
\lim_{l\to\infty}\lrangle{Z_1 Z_l}\to C\neq 0
\label{eq:z2_ssb_correlator}
\end{equation}

We now represent the $\ZZ_2$ even cat state using MPS, 
\begin{equation}
    \ket{\psi_e}=\sum_{\{j\}}\Tr[\cdots M_{j_l}\cdot M_{j_{l+1}} \cdots]\ket{\dots j_l j_{l+1}\dots}
\end{equation}
where $M_{j,\alpha\beta}$ is a three leg local tensor.
Imposing $\ZZ_2$ symmetry on the MPS requires $M$ satisfies the symmetry condition:
\begin{equation}
    \sum_{j'} X_{jj'}\cdot M_{j'}=W(g)\cdot M_j\cdot W^\dg(g) 
\end{equation}
where the gauge transformation $W(g)$ satisfies $[W(g)]^2=\hat{1}$.

Physical observables, such as those in Eq.~\eqref{eq:z2_ssb_correlator}, are calculated via the transfer matrix $T_{\alpha\alpha',\beta\beta'}\equiv \sum_{j}M_{j;\alpha\beta}\cdot M^*_{j;\alpha'\beta'}$, which inherits the $\ZZ_2$ symmetry generated by $W(g)\otimes W^*(g)$.
Assuming $T$ possesses a unique dominant eigenvalue~(normalized to $1$), with corresponding left and right eigenvectors $\vbra{v_l}$ and $\vket{v_r}$,  Eq.~\eqref{eq:z2_ssb_correlator} are then expressed as
\begin{equation}
    \lrangle{Z_1}=\vbra{v_l}T(Z)\vket{v_r}~,\quad
    \lim_{l\to\infty}\lrangle{Z_1 Z_l}=[\vbra{v_l}T(Z)\vket{v_r}]^2
\end{equation}
where $[T(Z)]_{\alpha\alpha',\beta\beta'}=\sum_{jj'}M_{j,\alpha\beta}\cdot Z_{jj'} M_{j';\alpha'\beta'}$.
Both terms vanish, since $T(Z)$ is $\ZZ_2$ odd while $\vbra{v_l}$ and $\vket{v_r}$ share the same $\ZZ_2$ quantum number.

To recover the long-range correlation of $\ZZ_2$-charged operators, a two-fold degeneracy in the dominant eigenvectors of $T$ with opposite $\ZZ_2$ quantum numbers is required. 
Denoting these (right) eigenstates as $\vket{v^e_{r}}$~($\ZZ_2$ even) and $\vket{v^o_{r}}$~($\ZZ_2$ odd),  Eq.~\eqref{eq:z2_ssb_correlator} become
\begin{equation}
\begin{aligned}
    \lrangle{Z_1} &= \frac{1}{2} \Big[\vbra{v^e_l}T(Z)\vket{v^e_r}+ \vbra{v^o_l} T(Z)\vket{v^o_r}\Big]=0\\
    \lim_{l\to\infty}\lrangle{Z_1 Z_l}&=\frac{1}{2}\cdot \Tr\Big[ (\vket{v^e_r}\vbra{v^e_l}+\vket{v^o_r}\vbra{v^o_l})\cdot T(Z)\cdot (\vket{v^e_r}\vbra{v^e_l}+\vket{v^o_r}\vbra{v^o_l})\cdot T(Z) \Big]\\
    &=\frac{1}{2}\Big[ \vbra{v^e_l}T(Z)\vket{v^o_r}\vbra{v^o_l}T(Z)\vket{v^e_r} + \vbra{v^o_l}T(Z)\vket{v^e_r}\vbra{v^e_l}T(Z)\vket{v^o_r} \Big]\sim O(1)
\end{aligned}
\label{eq:z2_ssb_order_param}
\end{equation}
where the factor $\frac{1}{2}$ comes from wavefunction normalization $\Tr(T^L)=\vbraket{v_l^e}{v_r^e}+\vbraket{v_l^o}{v_r^o}$. 

This degeneracy are most naturally enforced by introducing an additional $\ZZ_2$ symmetry acting exclusively on virtual legs, satisfying
\begin{equation}
    M_j=J\cdot M_j\cdot J^\dg~,\quad
    \text{where}~~J^2=\hat{1}~\text{and}~\{W(g),J\}=0
\end{equation}
Physically, $J$ can be viewed as the ``renormalized'' $\ZZ_2$ charge. 
The $\ZZ_2$ virtual symmetry imposes additional $\ZZ_2\times\ZZ_2$ symmetry on transfer matrix $T$, generated by $J\otimes \hat{1}$ and $\hat{1}\otimes J^*$, both anti-commuting with $W(g)\otimes W^*(g)$. 
Therefore, we have $\vket{v_r^o}=(J\otimes\hat{1})\vket{v_r^e}$.
The $\ZZ_2$ odd ground state of $\ZZ_2$ SSB phase can then be constructed by inserting $J$ on a virtual legs:
\begin{equation}
    \ket{\psi_o}=\sum_{\{j\}}\Tr[\cdots M_{j_l}\cdot J\cdot M_{j_{l+1}} \cdots]\ket{\dots j_l j_{l+1}\dots}
    \label{eq:z2_odd_state_from_igg}
\end{equation}
As an example, when $h=0$ in Eq.~\eqref{eq:tfim_ham}, the $\ZZ_2$ even ground state is $\ket{\psi_e}=\ket{\dots\uparrow\uparrow\dots}+\ket{\dots\downarrow\downarrow\dots}$, represented by an MPS with local tensor $M= \ket{\uparrow}\vket{\uparrow}\vbra{\uparrow} + \ket{\downarrow}\vket{\downarrow}\vbra{\downarrow}$.
Here, $W(g)=X$ and $J=Z$. 
$\ket{\psi_o}=\ket{\dots\uparrow\uparrow\dots}-\ket{\dots\downarrow\downarrow\dots}$ is obtained by Eq.~\eqref{eq:z2_odd_state_from_igg}.

We now calculate the $\ZZ_2$ domain wall correlator $\lrangle{\prod_{j=1}^l X_j}$ using transfer matrix: 
\begin{equation}
    \lrangle{\prod_{j=1}^l X_j}=\Tr\left[ (W(g)\otimes \hat{1})\cdot T^l\cdot (W^\dg(g)\otimes \hat{1}) \cdot T^{L-l}\right]\Big/\Tr(T^L)
    \label{eq:disorder_param_ising}
\end{equation}
For the $\ZZ_2$ symmetric phase ($h>J$), the transfer matrix has a unique dominant eigenvector, and Eq.~\eqref{eq:disorder_param_ising} reduces in the thermodynamic limit $L\gg l\to\infty$ to $\vbra{v_l}W(g)\vket{v_r}\cdot \vbra{v_l}W^\dg(g)\vket{v_r}$, giving a non-zero constant — the disorder parameter is long-range ordered.
For the $\ZZ_2$ SSB phase, the disorder parameter evaluates as
\begin{equation}
\begin{aligned}
    \lrangle{\prod_{j=1}^l X_j}
    =&\Bigg[\sum_{\alpha,\beta=e,o}\vbra{v_l^\alpha} W(g)\otimes \hat{1} \vket{v_r^\beta} (v_l^\beta| W^\dg(g)\otimes \hat{1} \vket{v_r^\alpha}\\
    +&\sum_{\alpha=e,o}\vbra{v_l^\alpha} {W(g)\otimes \hat{1}} \cdot \left( \sum_{a}\lambda_a^{l-1}\vket{w_{a,r}}\vbra{w_{a,l}} \right) \cdot {W^\dg(g)\otimes \hat{1}}\vket{v_r^\alpha}\Bigg]\Bigg/\Tr(T^L)
\end{aligned}
    \label{eq:disorder_param_ssb}
\end{equation}
where $\vket{w_{a,r}}/\vbra{w_{a,l}}$ are right/left eigenstates of $T$ with eigenvalue $\lambda_a$ satisfying $1>\abs{\lambda_1}\ge\lambda_{2}\ge\cdots$.
Crucially, symmetry $J\otimes J^*$ commutes with the $\ZZ_2$ symmetry $W(g)\otimes W^*(g)$, ensuring $\vket{v_r^e}$ and $\vket{v_r^o}$ share the same quantum number under $J\otimes J^*$.
However, as $W(g)\otimes\hat{1}$ is odd under $J\otimes J^*$, the first term in Eq.~\eqref{eq:disorder_param_ssb} vanishes.
We then conclude that for MPS with virtual leg symmetry $J$, domain wall correlator $\lrangle{\prod_{j=1}^l X_j}\sim\exp(-l/\xi)$, with $\xi=-(\ln\lambda_1)^{-1}$ (assuming $\vket{w_{1,r}}$ and $\vket{v_r^{e/o}}$ carry opposite quantum numbers under $J\otimes J^*$).

\subsection{1D mixed states with strong $\ZZ_2$ symmetry and MPDO}
Having reviewed the pure-state MPS formalism, we now turn to mixed states described by MPDOs. 
We start from the MPDO $\rho$
\begin{equation}
    \rho=\sum_{\{u,d\}}\Tr[\cdots M_{u_ld_l}\cdot M_{u_{l+1}d_{l+1}} \cdots]\ket{\dots u_l u_{l+1}\dots}\bra{\dots d_l d_{l+1}\dots}
\end{equation}
where $M_{ud,\alpha\beta}$ is a four-leg local tensor.

Now, we add strong $\ZZ_2$ symmetry $g=\prod_j X_j$, which imposes local constrains on $M$:
\begin{equation}
\begin{aligned}
    &\sum_{u'} X_{uu'}\cdot M_{u'd}=W_u(g)\cdot M_{ud}\cdot W_u^\dg(g) \\
    &\sum_{d'}  M_{ud'}\cdot X_{d'd}^\dg=W_d(g)\cdot M_{ud}\cdot W_d^\dg(g) \\
\end{aligned}
\end{equation}
where gauge transformation $W(g)$ satisfies $[W(g)]^2=\hat{1}$. 

As discussed in the main text, its different phases are characterized through the correlators of $\ZZ_2$-charged operators and $\ZZ_2$ domain walls. 
In particular, for the $\ZZ_2$ SW-SSB phase, it satisfies
\begin{equation}
   \lim_{l\to\infty}\Tr{(\rho \prod_{j=1}^l X_j)}\to  0~,\quad
\lim_{l\to\infty}\Tr{(\rho Z_1 Z_l)}\to  0
\label{eq:z2_swssb_correlator}
\end{equation}
Such observables in MPDO are calculated through the transfer matrix $T_{\alpha\beta}=\sum_u M_{uu,\alpha\beta}$, which holds a $\ZZ_2$ symmetry: 
\begin{equation}
    T = (W_u(g)W_d(g))T(W_u(g)W_d(g))^\dg
    \label{eq:mpdo_transfer_mat_z2_sym}
\end{equation}

Since $\rho$ is a short-range correlated, $T$ holds a non-degenerate leading eigenvector, with spectral decomposition $T=\vket{v_r}\vbra{v_l}+\cdots$, where the leading eigenvalue set to be $1$, and ``$\cdots$'' denotes contributions from other smaller eigenvalues.
Under $\ZZ_2$ symmetry $W_u(g)W_d(g)$ in Eq.~\eqref{eq:mpdo_transfer_mat_z2_sym}, $\ket{v_r}$ and $\bra{v_l}$ share the same quantum number.

In the thermodynamic limit, charge and domain-wall correlators are
\begin{equation}
    \begin{aligned}
         &\lim_{l\to\infty}\Tr{(\rho \prod_{j=1}^l X_j)} =\Tr[ T(X_1) T(X_2)\cdots T(X_l) TT\cdots] =\abs{\vbraopket{v_l}{W_u}{v_r}}^2 = \abs{\vbraopket{v_l}{W_d}{v_r}}^2 \\
         &\lim_{l\to\infty}\Tr{(\rho Z_1Z_l)} =\Tr[ T(Z_1) T\cdots T T(Z_l) TT\cdots] =\abs{\vbraopket{v_l}{T(Z)}{v_r}}^2 \\
    \end{aligned}
\end{equation}
where $[T(O)]_{\alpha\beta}\equiv\sum_{ud} O_{ud}M_{du,\alpha\beta}$.

As $T(Z)$ is odd under $W_u(g)W_d(g)$, the second line vanishes.
To make the first line vanishes, we introduce an additional $\ZZ_2$ virtual-leg symmetry $\Lambda$ on $M$:
\begin{equation}
   M_{ud,\alpha\beta} = \sum_{\alpha',\beta'} \Lambda_{\alpha\alpha'} M_{ud,\alpha'\beta'} \Lambda_{\beta'\beta}^{-1}~,\quad \Lambda^2 =\hat{1}
\end{equation}
which leads to a new $\ZZ_2$ symmetry of $T$: 
\begin{equation}
    T=\Lambda\cdot T \cdot\Lambda^{-1}
    \label{}
\end{equation}
By further imposing $\{\Lambda,W_{u/d}\}=0$, $\vbraopket{v_l}{W_{u/d}}{v_r}=0$ due to symmetry constraints.

To summarize, the MPDO description of the $\ZZ_2$ SW-SSB phase is captured by the following tensor network conditions:
\begin{equation}
    \rho=
    \adjincludegraphics[scale=1,valign=c]{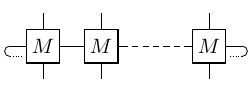}
    \label{}
\end{equation}
where 
\begin{equation}
\begin{aligned}
\adjincludegraphics[scale=1,valign=c]{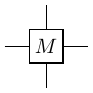} 
   & =
    \adjincludegraphics[scale=1,valign=c]{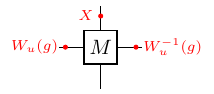} 
    =
    \adjincludegraphics[scale=1,valign=c]{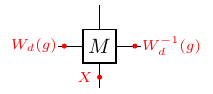}  =
     \adjincludegraphics[scale=1,valign=c]{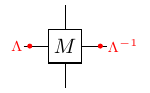} \\
\end{aligned}
    \label{}
\end{equation}
with $\{\Lambda,W_{u/d}\}=0$.

\section{Disorder parameter for $\ZZ_2$ SET phase}\label{app:z2_set_imo}
The pure--mixed correspondence established in the main text maps a 2D topological phase to a 1D mixed state $\rho_{\partial A}$.
This mapping motivates new observables for gappedd $\ZZ_2$ spin liquid phases, where $e$-anyon carries spin-$\frac{1}{2}$. 
In this section, we propose a new diagnostic for $\ZZ_2$ symmetry fractionalization in a $\ZZ_2$ toric code SET: the long-range $m$-anyon correlator in the presence of a $\ZZ_2$ disorder operator.

We consider a $\ZZ_2$ topological order enriched by a global $\ZZ_2$ symmetry $G = \{1, g\}$.
We focus on the pattern where the $e$-anyon carries a projective (half-integer) representation of $G$:
\begin{equation}
    U_e^2(g) = -1,
\end{equation}
where $U_e(g)$ is the action of $g$ in the vicinity of a single $e$-anyon.
Physically, this means that fusing two $g$-symmetry defects produces an $m$-particle.
Our goal is to detect this fractionalization pattern from long-range behavior of some physical observables. 

\begin{figure}[htpb]
    \centering
\adjincludegraphics[scale=1,valign=c]{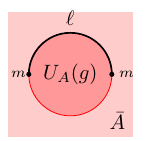}
    \caption{Schematic of the ``interspersed membrane'' operator: the disorder operator $U_A(g)$ (red shaded region $A$) is interspersed with two $m$-anyons (black dots) separated by distance $\ell$.}
    \label{fig:interspersed_membrane}
\end{figure}

The key idea is to evaluate the $m$-anyon correlator in the presence of a symmetry-twist (disorder operator) $U_A(g)$ on a subregion $A$.
We propose the normalized correlator
\begin{equation}
    \lrangle{m_im_j}_g\equiv\frac{\lrangle{m_i m_j U_A(g)}}{\lrangle{U_A(g)}}\xrightarrow{|i-j|\rightarrow\infty}\begin{cases}
                O(1) &{\rm{for}}~ U_e^2(g)=-1, \\
                0    &{\rm{for}}~U_e^2(g)=1,
            \end{cases}
    \label{eq:cor_sf}
\end{equation}
    for $i$ and $j$ near the entanglement cut $\partial A$, as illustrated schematically in Fig.~\ref{fig:interspersed_membrane}.

In the following, we will prove Eq.~\eqref{eq:cor_sf} by the MPDO formalism: the 1D RDM $\rho_{\partial A}$ is obtained by contracting internal legs of tensor $M$, where the SET structure imposes the following constraints on $M$:
\begin{equation}
    \adjincludegraphics[scale=1,valign=c]{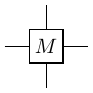} 
    =
    \adjincludegraphics[scale=1,valign=c]{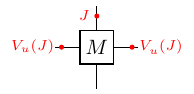} 
    =
    \adjincludegraphics[scale=1,valign=c]{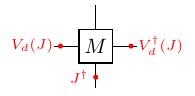} 
    =
    \adjincludegraphics[scale=1,valign=c]{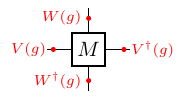} 
    =
    \adjincludegraphics[scale=1,valign=c]{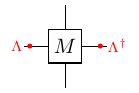},
    \label{eq:teneq_z2_strong_sym_SM}
\end{equation}
Here, the first and second tensor equations enforce the strong $\ZZ_2$ symmetry generated by $J$, originating from the $\ZZ_2$ topological order.
The third equation encodes the weak symmetry generated by $W(g)$, inherited from the global $\ZZ_2$ symmetry. 
The fourth equation, involving $\Lambda$ with $\{\Lambda, V_{u/d}(J)\} = 0$, ensures that $\rho_{\partial A}$ realizes the SW-SSB phase, as established in Sec.~\ref{app:mps_mpdo_correlator}.
    Further, the fractionalization condition $U_e(g)^2 = -1$ translates to
\begin{equation}
    W(g)^2 = J.
\end{equation}
Physically, this means that fusing two $g$-defect produces an $m$-particle — the hallmark of the nontrivial SET.

Another ingredient needed for the proof is the Hermiticity of $\rho_{\partial A}$, which imposes the following constraint on $M$:
    \begin{equation}
        \begin{aligned}
            \adjincludegraphics[scale=1,valign=c]{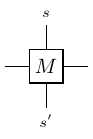}
            =
            \adjincludegraphics[scale=1,valign=c]{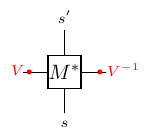},
        \end{aligned}
        \label{eq:hermit}
    \end{equation}
    where $V_\HH\cdot V_\HH^*=\hat{1}$, and $s$ and $s'$ label the virtual bond indices of the bra and ket layers.

To evaluate the correlator $\lrangle{m_i m_j}_g$, we examine the transfer matrix with a $W(g)$ insertion (the symmetry-twisted transfer matrix):
    \begin{equation}
    T(g) \equiv  \adjincludegraphics[scale=1,valign=c]{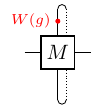} 
\end{equation}
Symmetries of $T(g)$ can be derived from those of $M$. 
In particular, 
\begin{equation}
    \begin{aligned}
        [T(g)]^{*} &=
        \adjincludegraphics[scale=1,valign=c]{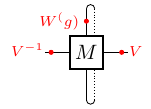}
        = 
        \adjincludegraphics[scale=1,valign=c]{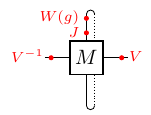}
        &= V_\HH^{-1} V_{u}(J)^{-1}\cdot \, T(g) \,\cdot V_{u}(J) V_\HH,
    \end{aligned}
\end{equation}
where Eq.~\eqref{eq:hermit} is used for the first equality, $W(g)^2 = J$ is used in the second equality, and in the third equality the strong-symmetry tensor condition (Eq.~\eqref{eq:teneq_z2_strong_sym_SM}, first two relations) was used to push $J$ through $T$.
Thus $T(g)$ is invariant under the combined operation $V_{u}(J) V_\HH \mathcal{K}$, where $\mathcal{K}$ denotes complex conjugation.

The transfer matrix also inherits the virtual $\ZZ_2$ symmetry: $[\Lambda, T(g)] = 0$.
Crucially, due to the anti-commutation relation $\{\Lambda, V_{u}(J) V_\HH \mathcal{K}\} = 0$, the leading eigenstate of $T(g)$ is at least two-fold degenerate (in terms of modulus).
In the $\Lambda$-diagonal basis, the two dominant (right) eigenstates, denoted as $\ket{r_1}$ and $\ket{r_2}=V_{u}(J) V_\HH \mathcal{K}\ket{r_1}$, hold distinct quantum number under $\Lambda$.

As discussed in supplemental section~\ref{app:mps_mpdo_correlator}, such degeneracy can be detected by long-range correlators of $V_u(J)$ on internal legs, which is odd under $\Lambda$, leading to $\vbra{l_1} V_u(J) \vket{r_1} = 0$.
Physically, such correlator is interpreted as correlator of $m$-anyons in the presence of the disorder operator.
Namely,
\begin{equation}
    \begin{aligned}
        \lrangle{m_im_j}_g&=\frac{\Tr[m_i m_j W_{\partial A}(g)\rho_{\partial A}]}{\Tr[W_{\partial A}(g)\rho_{\partial A}]}\\
        &\rightarrow \sum_{a,b=1}^2 \alpha^\ell \alpha^{*L-\ell}\frac{\vbra{l_a}V_{u}(J)\vket{r_b}\vbra{l_b}V_{u}(J)^\dagger\vket{r_a}}{\alpha^L+\alpha^{*L}}\\
        &\sim O(1)
    \end{aligned}
\end{equation}
where $\alpha$ and $\alpha^*$ denote two leading eigenvalues of $T(g)$.

\paragraph{PEPS construction for the $\ZZ_2$ SET model.}
To verify the proposed diagnostic, we perform numerical simulations on a $\ZZ_2$ SET phase.
Below we construct the explicit PEPS tensors for the $\ZZ_2$ enriched toric code model by ungauging the $\ZZ_4$ gauge theory~\cite{set1Heinrich2016,set2Cheng2017, Iqbal2018}.
The physical leg is defined on the edge, with dimension four. 
We define the physical operators $\tilde X$ and $\tilde Z$ as
\begin{equation}
    \tilde X = \left(
    \begin{array}{cccc}
     & 1 & & \\
     &  & 1 & \\
     &  & & 1 \\
     1 & & & 
    \end{array}
    \right), \tilde Z =  \left(
    \begin{array}{cccc}
       1 & & &\\
       & i &  &\\
       & & -1 &\\
       & & & -i
    \end{array}
    \right)
\end{equation}
satisfying  $\tilde X\tilde Z= i\tilde Z\tilde X$. We also introduce a spin-1/2 on each vertex, whose Pauli operators are denoted as $\tau^{x,y,z}$. The PEPS formalism consists site tensor $V$ and bond tensor $E$ both with bond dimension 4, such that the following constraints hold:
\begin{equation}
     V = \adjincludegraphics[scale=1,valign=c]{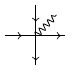}
= \adjincludegraphics[scale=1,valign=c]{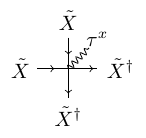}
=\adjincludegraphics[scale=1,valign=c]{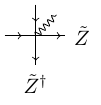}
=
\adjincludegraphics[scale=1,valign=c]{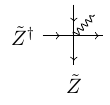}
= \adjincludegraphics[scale=1,valign=c]{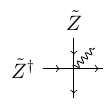},\label{V_tensoreq}
\end{equation}
\begin{equation}
   E =\adjincludegraphics[scale=1,valign=c]{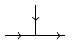}
    =\adjincludegraphics[scale=1,valign=c]{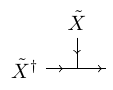}
    =\adjincludegraphics[scale=1,valign=c]{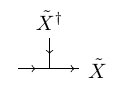}
    =\adjincludegraphics[scale=1,valign=c]{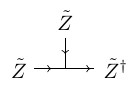},
\end{equation}
where an inward(outward) arrow denotes a local Hilbert space for ket(bra).
Here, the $e$ anyon is identified as $\tilde{Z}$ or $\tilde{Z}^3$ on a virtual leg, while $m$-anyon is identified as the end point of $\tilde{X}^2$ string on virtual legs.
A global $\ZZ_2$ symmetry $U(g)=\prod_j \tau_j^x$ is also imposed by the second equation in Eq.~\eqref{V_tensoreq}, where the virtual leg action $W(g)=\tilde{X}$.
Here $W(g)^2=\tilde{X}^2$ is interpreted as fusion of two $g$-defects leads to an $m$-anyon.
Therefore, the resulting phase is the $\ZZ_2$ symmetry enriched $\ZZ_2$ toric code topological order, where $g^2\circ e=-e$.

\begin{figure}
        \centering
        \includegraphics[width=0.6\linewidth]{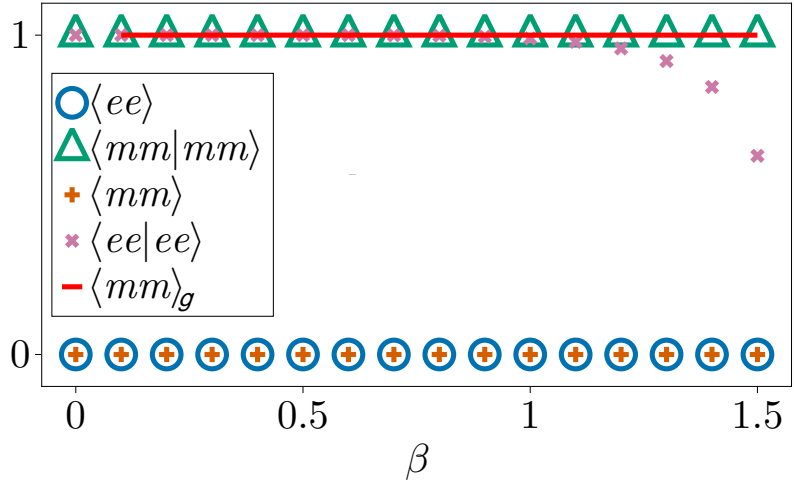}
        \caption{Numerical result for $\lrangle{mm}_g$ as a diagnostic of $\ZZ_2$-symmetry fractionalization.
        The model is deformed toric code wave function enriched by a global $\ZZ_2$ symmetry.
        $\lrangle{mm}_g$ is non-vanishing in the thermodynamic limit for the case where $e$-anyon carries fractional $\ZZ_2$ quantum number.
        Here, cut-off bond dimension $\chi=30$ to obtaining boundary MPS.
        }
        \label{fig:z2frac_toric_code}
    \end{figure}

The PEPS could be perturbed by applying a local filtering operator  $F  = \exp\{\frac{1}{4}(\beta_x X+\beta_x X^\dagger+\beta_z Z+\beta_z Z^\dagger)\}$ on each bond, generating a family of states.
Here we take $\beta_x=\beta_z=\beta$.
We then evaluate the twisted correlator $\lrangle{m_i m_j}_g$ is evaluated in the thermodynamic limit with cut off bond dimension $\chi=30$, where boundary MPS are obtained by the VUMPS method.
As shown in Fig.~\ref{fig:z2frac_toric_code}, $\lrangle{m_i m_j}_g$ approaches a non-zero constant at large separation when the $\ZZ_2$ symmetry is fractionalized ($U_e(g)^2 = -1$), confirming the analytical prediction.
Note that when $\beta = 0$, $\lrangle{U_A(g)}$ vanishes, and the normalized correlator is recovered via the limiting procedure $\lim_{\beta \to 0} \lrangle{m_i m_j}_g$.

\end{document}